\pdfoutput=1
\documentclass[12pt]{article}

\usepackage{amsmath,amssymb,bm}
\usepackage{graphicx}
\usepackage{booktabs}
\usepackage{placeins}
\usepackage{microtype}
\usepackage{float}
\usepackage{pgfplots}
\pgfplotsset{compat=1.18}
\usepgfplotslibrary{groupplots}

\usepackage{braket}
\usepackage{geometry}
\usepackage[
colorlinks=true,
linkcolor=blue,
citecolor=blue,
urlcolor=blue,
breaklinks=true
]{hyperref}

\title{\textbf{Separable representations of two-body interactions for Faddeev calculations of light nuclei}}
\author{M.~M.~Nishonov\thanks{Email: m.nishonov@nuu.uz}\\
	National University of Uzbekistan, 100174 Tashkent, Uzbekistan}
\date{}

\begin{document}
\maketitle

\begin{abstract}
Separable representations of the Ernst--Shakin--Thaler type are constructed for the charge-dependent Bonn (CD~Bonn) nucleon--nucleon potential ($j\le4$; $nn$, $np$, $pp$), for the Kanada--Kaneko--Nagata--Nomoto (KKNN) $\alpha N$ interaction, and for the Buck--Friedrich--Wheatley (BFW) $\alpha\alpha$ interaction. They are the input for momentum-space Faddeev calculations of $A=3$, $6$, $9$, and $12$ systems. The representations are tested against six acceptance criteria: off-shell accuracy at the pair energies sampled by the three-body kernel, phase shifts, effective-range parameters, bound and resonant states, and numerical stability.
\end{abstract}

\section{Introduction}
\label{sec:intro}

Momentum-space Faddeev calculations with separable two-body input remain an economical route to the bound states and low-energy scattering of light three-body systems, from the three-nucleon system to the light nuclei which the cluster model describes as three bodies, $^{6}\mathrm{He}$ and $^{6}\mathrm{Li}$ as $\alpha+N+N$, $^{9}\mathrm{Be}$ and $^{9}\mathrm{B}$ as $\alpha+\alpha+N$ and $^{12}\mathrm{C}$ as $\alpha+\alpha+\alpha$, with the nucleon--nucleon, $\alpha N$ and $\alpha\alpha$ pair interactions as the only input; three-body models of $^{6}$He and $^{6}$Li with separable cluster interactions go back to Ghovanlou and Lehman~\cite{Ghovanlou1974} and to Eskandarian and Afnan~\cite{Eskandarian1992}, and separable representations of the same Ernst--Shakin--Thaler (EST) type~\cite{EST1973} were built for nucleon--nucleus optical potentials, with the Coulomb force carried in the Coulomb-distorted basis~\cite{Hlophe2013,ElsterNTSE2014}. The formal limitation is well known. A finite-rank representation is exact only at its support energies, the energies at which it coincides with the exact two-body amplitude, and the two-body scattering data leave the interaction free off shell (Ekstein~\cite{Ekstein1960}); Polyzou and Gl\"ockle~\cite{PolyzouGloeckle1990} showed that within a single nucleus this residual off-shell freedom is indistinguishable from a three-body force. To interpret the residual $E_{\mathrm{exp}}-E_{2\mathrm{BF}}$, the difference between the experimental energy and the energy computed from two-body forces alone, as three-body physics, a calculation must therefore quantify two things that are usually mixed together: how well the original potentials, the potentials being represented, were reproduced, and what the finite-rank approximation costs. The first source of error is the reproduction of the original potential itself.  The second follows from the choice of the support points and of the rank, the number of separable terms. The most common practice is to compare only the separable potentials with experiment, and this makes the two indistinguishable.

This paper provides that quantification for the two-body input of a unified set of Faddeev calculations covering $A=3$, $6$, $9$, and $12$: the charge-dependent Bonn potential (CD~Bonn)~\cite{Machleidt2001} in all partial waves with $j\le4$ and all three charge states; the Kanada--Kaneko--Nagata--Nomoto (KKNN) $\alpha N$ interaction~\cite{KKNN1979}; and the Buck--Friedrich--Wheatley (BFW) $\alpha\alpha$ interaction~\cite{BFW1977}.

These four mass numbers reduce to three constituents interacting through these pairs and are therefore solvable exactly from the two-body input by the Faddeev equations, with no free many-body parameter in the input. For $A=6$, $9$ and $12$ the constituents are $\alpha$~clusters and valence nucleons: the $\alpha$ particle, bound by $28.3$~MeV with its first excited state near $20$~MeV, stays in its ground state at the energies of these systems and acts as an inert, structureless boson, and the KKNN and BFW potentials are the effective $\alpha N$ and $\alpha\alpha$ interactions of that picture. $A=3$ ($^{3}$H, $^{3}$He) tests the $NN$ force alone, $A=6$ ($^{6}$He, $^{6}$Li) combines $NN$ with $\alpha N$, $A=9$ ($^{9}$Be) combines $\alpha N$ with $\alpha\alpha$, and $A=12$ ($^{12}$C) tests $\alpha\alpha$ alone; through the charged pairs $pp$, $\alpha p$ and $\alpha\alpha$ they also test the Coulomb-dependent content of the input. The three-body tests of the present paper are those at $A=3$ and $6$; the $A=9$ and $12$ systems, in which the $\alpha\alpha$ Coulomb content dominates, are treated in Ref.~\cite{FockII} on the same input.

The separable representations are of the EST type, in the tradition begun for realistic nucleon--nucleon ($NN$) potentials by Haidenbauer and Plessas~\cite{Haidenbauer1984} and continued for optical and cluster interactions in Refs.~\cite{Hlophe2013,Hlophe2017}. The coupling matrix is rebuilt from analytically fitted form factors, so the separable potentials entering the three-body kernel (the integral operator of the momentum-space Faddeev equations~\cite{Nishonov2026a}) are internally consistent. Published separable representations of individual interactions are not taken over, because a three-body calculation of these systems needs one consistent set across all pairs in a single analytic form-factor family, with the Coulomb-dressed Pauli-forbidden eigenstates of each interaction, that is, the eigenstates computed with the Coulomb force included where it acts, carried as part of the representation, and with the off-shell deviation from the original potential quantified on the domain the kernel samples, which is the subject of this paper. The same idea is returning in modern form in the low-rank factorizations of chiral interactions~\cite{Tichai2019}; the properties tested below (residue structure, threshold behavior, coupled-channel sign conventions) concern any low-rank representation, whatever the original interaction, and the criteria and the error budget they yield should apply to chiral and pionless separable interactions as well.

Beyond the separable potentials themselves, the paper states the tests by which they are accepted. Six two-body criteria (Sec.~\ref{sec:criteria}) compare each representation with its original potential: the off-shell $t$~matrix at the pair energies the three-body kernel samples, the phase shifts between the supports, the effective-range parameters and the asymptotic $D/S$ ratio of the deuteron recomputed from the fitted form factors, the bound states, the resonances, and the numerical stability of the tabulated coefficients. One condition lies beyond them: the sign of the off-diagonal component of a coupled channel relative to the recoupling convention of the three-body kernel, that is, the angular-momentum coupling conventions under which the kernel combines the pair channels, is not fixed by any two-body criterion: a comparison with the original potential fixes the sign relative to that potential, but the convention of the original potential relative to the recoupling of the kernel is tested only by a three-body observable; it is fixed by the neutron--deuteron $^{4}P_{J}$ eigenphases (the phase shifts of the diagonalized multichannel $S$~matrix), which depend on it at first order (Sec.~\ref{sec:dsign}). The three-body calculations of Sec.~\ref{sec:threebody} give the size of the effects which the criteria and this condition constrain: the asymptotic $D/S$ ratio, a ratio of residues at the deuteron pole which neither the binding energy nor the phase shifts at the supports fix, is worth about $90$~keV in the triton and in $^{6}$Li for a $3.4\%$ deviation at unchanged rank (Sec.~\ref{sec:eta}); the threshold of the $^{1}S_{0}$ $nn$ potential, which supports at the negative energies of the kernel alone do not protect, is worth $15$~keV in $^{6}$He for a $10\%$ deviation of the scattering length (Sec.~\ref{sec:1s0}); the sign of the $^{3}S_{1}$--$^{3}D_{1}$ off-diagonal component is worth $7$~keV in the triton with all $j\le4$ waves, but degrees in the $nd$ $^{4}P_{J}$ eigenphases. A phenomenological potential must further be used together with the Coulomb interaction it was fitted with: replacing the $\alpha\alpha$ Coulomb interaction of the BFW model, folded over the charge distributions of the two $\alpha$ particles, by the point form displaces the $^{8}$Be ground-state resonance by $0.36$~MeV and the Pauli-forbidden states by MeV (Sec.~\ref{sec:folded}). The validation of two-body separable potentials therefore cannot stop at the two-body level.

A brief summary of the three-body framework, the projection of Pauli-forbidden states and the treatments of the Coulomb interaction opens the paper (Sec.~\ref{sec:methods}); the acceptance criteria are formulated in Sec.~\ref{sec:criteria}, and the construction of the separable potentials is summarized in Sec.~\ref{sec:construction}. The three interactions occupy Secs.~\ref{sec:nn}--\ref{sec:aa}, each with the three-way comparison of separable potential, original potential and experiment that the criteria require. The consequences at the three-body level are examined in Sec.~\ref{sec:threebody}, and Sec.~\ref{sec:summary} collects the results. The projection of Pauli-forbidden states follows the Feshbach--Schur construction of Ref.~\cite{Nishonov2026a}. The treatments of the $\alpha p$ and $\alpha\alpha$ Coulomb interaction in the three-body equations are described in Sec.~\ref{sec:coulomb} of the present paper. The separable potentials themselves are purely nuclear. Their only Coulomb-dependent content is the forbidden states they carry, which are the Coulomb-dressed ones.

\section{Framework and methods}
\label{sec:methods}

The validation criteria of this paper refer to the three-body framework the separable potentials are intended for. The equations of that framework are derived in Ref.~\cite{Nishonov2026a}. This section describes only what the criteria depend on: where the separable input enters the three-body kernel, how Pauli-forbidden states are projected out, and the three treatments of the Coulomb interaction used in this work.

\subsection{Separable input in the three-body kernel}
\label{sec:ags}

Bound states are computed from the homogeneous form of the momentum-space Faddeev equations of Alt, Grassberger and Sandhas~\cite{AGS1967}, as reformulated for three particles of arbitrary identity in Ref.~\cite{Nishonov2026a}, in the line of the three-cluster bound-state calculations of Ref.~\cite{Blokhintsev2006PAN}. The neutron--deuteron eigenphases of this paper are computed from the inhomogeneous (scattering) form of the same equations, in the same way as in earlier proton--deuteron scattering calculations~\cite{Alt2002}. Every pair interaction is a separable potential,
\begin{equation}
V=\sum_{ij}|\chi_{i}\rangle\Lambda_{ij}\langle \chi_{j}|,
\label{eq:seppot}
\end{equation}
whose two-body $t$~matrix takes the same separable form,
\begin{equation}
t(E)=\sum_{ij}|\chi_{i}\rangle\tau_{ij}(E)\langle \chi_{j}|,
\label{eq:sepform}
\end{equation}
with $i,j=1,\dots,N_{r}$ and $N_{r}$ the rank, $\chi_{i}$ the form factors, $\bm{\Lambda}$ the coupling matrix and $\bm{\tau}(E)$ the pair $t$~matrix in the space of the form factors, all constructed in Sec.~\ref{sec:construction}. The equations then close on one-variable spectator amplitudes; particle identity is carried by multiplicity weights and exchange phases, and all dynamics, including the projection of the forbidden states, resides in the $\bm{\tau}$~matrix. The kernel evaluates $\bm{\tau}$ at the pair energy $E_{3}-q^{2}/2\bar{\mu}$ ($E_{3}$ the three-body energy, $q$ the spectator momentum, $\bar{\mu}$ the spectator reduced mass), which for a bound state is negative throughout: the entire two-body input enters the bound-state problem at negative pair energies, the region of criterion C1. In a Borromean system (a bound three-body system none of whose two-body subsystems is bound) the kernel reaches the near-threshold region of an unbound pair by analytic continuation from below threshold, the region of criterion C5. The three-body energies quoted in this paper are obtained by solving these equations on fixed Gauss--Legendre meshes.

\subsection{Pauli-forbidden states and their projection}
\label{sec:fsp}

Deep cluster potentials such as KKNN ($\alpha N$) and BFW ($\alpha\alpha$) simulate the effect of antisymmetrization by supporting deeply bound $S$- and $D$-wave states that are forbidden by the Pauli principle in the compound system. In a three-body calculation these states must be projected out.  The strict projection is used here. The orthogonalizing pseudopotential of Krasnopol'skii and Kukulin~\cite{Kukulin1974} adds the term
\begin{equation}
\lambda\sum_{f}|\phi_{f}\rangle\langle\phi_{f}|,
\label{eq:opp}
\end{equation}
the sum running over the forbidden states $\phi_{f}$ of the pair ($f$ labelling the state, its wave included), which pushes them to high energy for large coupling $\lambda$. The exact $\lambda\to\infty$ limit is taken algebraically, as a Schur complement on the forbidden block of the extended separable basis; the operator identity behind this construction (the Feshbach--Schur projection, in the sense of Feshbach's projection formalism~\cite{Feshbach1958,Feshbach1962}) is derived in Ref.~\cite{Nishonov2026a}. The elimination is the exact $\lambda\to\infty$ limit for any projected vector; it removes exactly the forbidden bound state and leaves the continuum untouched only if that vector is an eigenstate of the interaction actually in use~\cite{Nishonov2026a}. For a separable potential this means the forbidden state must be the bound state of the fitted representation itself, not of the original potential.  Every separable potential with a forbidden state carries it in this form, and criterion C4 verifies both count and position.

\subsection{Three Coulomb treatments}
\label{sec:coulomb}

The Coulomb interaction enters this work in three forms with different domains of validity.

\emph{(i) Screened Coulomb (two-body continuum only).} For two-body  \emph{continuum} observables (the $^{5}$Li and $^{8}$Be resonances) the point Coulomb is screened and the observables are computed at several screening radii ($R=50$--$200$~fm), with the $R$~dependence followed.  Where it has not died out at the largest radius the value is extrapolated in $R$ and the residual $R$~sensitivity is stated. The $^{5}$Li values use the Yukawa screening
\begin{equation}
V_{C}(r)\to Z_{1}Z_{2}e^{2}\,\frac{e^{-r/R}}{r},
\label{eq:yukawa}
\end{equation}
which has analytic partial-wave matrix elements and whose screening limit, after renormalization, restores the free-wave asymptotics~\cite{Taylor1974}. The $^{8}$Be values use the sharper screening $\exp[-(r/R)^{4}]$ of Ref.~\cite{Deltuva2005}, which converges faster in $R$.

\emph{(ii) Fock representation (separable representation of the unscreened Coulomb interaction).} In the bound-state three-body equations the Coulomb interaction of a charged pair is represented by a finite block of the Fock modes of Ref.~\cite{FockI}, an identity for the potential at negative energies. This expansion converges at the negative pair energies the kernel samples. The construction, its convergence, and its limitations in the continuum are validated in Ref.~\cite{FockI} against the exact Coulomb-modified phases and the screening--renormalization method, and the three-body bound states with that input are given in Ref.~\cite{FockII}; those papers take their nuclear separable input from the present work, so they are complementary, this one validating the nuclear input and Refs.~\cite{FockII,FockI} the Coulomb representation.  Here only the role of that representation as the source of the point-Coulomb interaction matters. Whenever a separable potential is said to be used together with the point Coulomb, this representation supplies the point-Coulomb interaction in the three-body equations. Coulomb-dressed forbidden states are computed in the same extended basis.

\emph{(iii) Folded Coulomb of an extended charge distribution.}  The finite charge distribution of the $\alpha$ particle folds the pair Coulomb interaction into
\begin{equation}
Z_{1}Z_{2}e^{2}\,\frac{\mathrm{erf}(\beta_{f}r)}{r},
\label{eq:folded}
\end{equation}
where the range parameter $\beta_{f}$ is set by the charge form factor. For the $\alpha\alpha$ system the BFW potential is defined together with such a folding, and the consequences are examined with the $\alpha\alpha$ interaction. The separable potentials are constructed so that the short-range separable potential plus the point-Coulomb representation of item~(ii) reproduces the folded interaction, with the accuracy given in Ref.~\cite{FockI}.

\section{The acceptance criteria}
\label{sec:criteria}

Two comparisons with different meanings run through this paper. The comparison of an \emph{original} potential with experiment quantifies the quality of the model.  Where the model was fitted to the same data, it only verifies that the original potential has been reproduced. The comparison of a \emph{separable potential} with its original potential quantifies the cost of the finite-rank approximation. Only the second is under the control of the present work, and only the first is fixed by the literature.  Their mixture (separable versus experiment) is therefore not used here on its own to judge a representation. It conflates the two effects, and is meaningful only read together with the comparison with the original potential, which the tables give wherever the original and the separable values differ. Note also that a separable potential which reproduces experiment better than its original potential does so by accidental cancellation, not by accuracy.

A separable potential is accepted when the following hold, C2 being reported as a diagnostic.

\paragraph*{C1: off-shell accuracy.}
The kernel of the bound-state Faddeev equations evaluates the two-body $t$~matrix at the pair energies $E=E_{3}-q^{2}/2\bar{\mu}$ introduced above, which extend to $-200$~MeV and below. The quantity to be reported is therefore
\begin{equation}
\Xi(E)=
\frac{\lVert t_{\mathrm{sep}}(p,p';E)-t_{\mathrm{orig}}(p,p';E)\rVert_{w}}
     {\lVert t_{\mathrm{orig}}(p,p';E)\rVert_{w}},
\label{eq:eps}
\end{equation}
where $p,p'$ run over the quadrature points of the kernel mesh up to $3\,\mathrm{fm}^{-1}$ and $E=-5$ to $-200$~MeV. The norm $\lVert\cdot\rVert_{w}$ is the mesh norm in which each point carries the weight $w=p^{2}w_{p}$ that the Faddeev kernel itself applies ($w_{p}$ are the quadrature weights). On-shell agreement does not constrain $\Xi$, and absolute thresholds for it proved to be interaction dependent. The criterion is therefore \emph{relative}: within a channel class the representation with the smaller $\Xi$ is retained, and the achieved values are reported per channel with the adopted set.

\paragraph*{C2: phases between the support points.}
EST separable potentials are exact at their supports by construction. The informative comparison is on a fine grid between them. The positive-energy phases are fixed by the positive half-shell supports, which is why the scattering region is reproduced despite the negative-energy supports that carry the off-shell content. C2 is a diagnostic rather than a requirement. Because the kernel samples off shell, a separable potential failing C2 while passing C1 is acceptable, and the reverse is not.

\paragraph*{C3: effective-range parameters and the asymptotic $D/S$ ratio from the fitted separable potentials.}
The scattering length $a$, the effective range $r$ and, for coupled channels, the asymptotic $D/S$ ratio of the bound-state wave function,
\begin{equation*}
\eta=\frac{A_{D}}{A_{S}},\qquad
u(r)\to A_{S}\,e^{-\gamma r},\qquad
w(r)\to A_{D}\,e^{-\gamma r}\Bigl[1+\frac{3}{\gamma r}+\frac{3}{(\gamma r)^{2}}\Bigr],
\end{equation*}
with $A_{S}$ and $A_{D}$ the asymptotic normalization constants of its $S$- and $D$-wave components $u$ and $w$ and $\gamma$ the bound-state wave number, are recomputed from the fitted analytic form factors, not from the exact half-shell ones from which the fit started. Pass: $a$ within $1\%$, $r$ within $2\%$, $\eta$ within $2\%$ of the original potential. Among the two-body criteria, the scattering length and the $D/S$ ratio are fixed by C3 alone: neither the phase shifts between the supports nor the binding energy fix them (Secs.~\ref{sec:eta} and~\ref{sec:1s0}). 

\paragraph*{C4: bound states.}
The number of bound states must match the original potential exactly and each energy to $5$~keV. The count is discrete, a sharper test than an energy, and for the $\alpha\alpha$ system it is also a Coulomb test. Two documented exceptions to the energy tolerance (the two $L=0$ $\alpha\alpha$ forbidden states) are discussed in Sec.~\ref{sec:aa}.

\paragraph*{C5: resonances from the fitted separable potentials.}
For channels whose resonances the three-body kernel samples, the fitted separable potential must reproduce the original potential's resonance position and width. The pole is not among the fit constraints, so its reproduction is a test. Section~\ref{sec:aa} reports one structural limitation in this respect.

\paragraph*{C6: numerical stability.}
$\max|a_{in}|\lesssim10^{3}$ for the tabulated Legendre coefficients of the analytic form-factor ansatz, Eq.~(\ref{eq:ffana}) below (beyond this the Legendre sums suffer severe cancellation); the asymmetry of the coupling matrix as rebuilt from the fitted factors, before it is symmetrized, at the level of the fit residual, $\sim10^{-4}$ (a rebuilt matrix symmetric to machine precision would indicate that it was not formed from the fitted factors, which produces an internally inconsistent pair); and fit residuals below $10^{-2}$.

Not every criterion constrains every interaction.  C1, C2, C4 and C6 apply throughout; the $\eta$ part of C3 only to the coupled $NN$ channel, single-channel $\alpha N$ and $\alpha\alpha$ having no such ratio; and C5 only to the $\alpha N$ and $\alpha\alpha$ interactions, whose subsystems have the resonances the $NN$ waves do not.

\section{Construction of the separable potentials}
\label{sec:construction}

All representations are of EST type~\cite{EST1973}: for support states $|\psi_{i}\rangle$ at energies $E_{i}$, $i=1,\dots,N_{r}$ with $N_{r}$ the rank of the representation,
\begin{equation}
\chi_{i}=V|\psi_{i}\rangle,\qquad
\bigl[\Lambda^{-1}\bigr]_{ij}=\langle\psi_{i}|V|\psi_{j}\rangle,
\label{eq:est}
\end{equation}
so that the representation
\begin{equation}
t_{\mathrm{sep}}(E)=\sum_{i,j=1}^{N_{r}}|\chi_{i}\rangle\tau_{ij}(E)\langle \chi_{j}|,
\qquad
\bm{\tau}^{-1}(E)=\bm{\Lambda}^{-1}-\bm{\mathcal D}(E),
\qquad
\mathcal D_{ij}(E)=\langle \chi_{i}|G_{0}(E)|\chi_{j}\rangle,
\label{eq:tsep}
\end{equation}
is exact half-on-shell (one momentum argument on the energy shell) at every support; $G_{0}(E)=(E-H_{0})^{-1}$ is the free resolvent. Support energies are chosen against criterion C1 rather than against phases, and every bound state of the original potential is included as a support. The remaining supports are half-shell columns of the standing-wave (principal-value) $K$~matrix, $\chi_{i}(p)=K(p,k_{i};E_{i})$, which coincide with the half-shell $t$~matrix on shell: at positive energies, where they suppress the spurious deep states that an all-negative support set can produce, and at negative energies, where the $K$ and $t$ matrices are the same real function, with $k_{i}$ a chosen off-shell momentum, for the off-shell region the kernel samples, the extension of the EST construction in which the energy and the momentum of a support are chosen independently~\cite{Hlophe2017}; for channels with long-range Coulomb structure the negative-energy supports also replace the scattering supports, since Coulomb-distorted scattering states are not smooth.

To use the separable potentials in the three-body equations, the exact half-shell form factors are fitted to the analytic form
\begin{equation}
\chi^{(L)}_{i}(p)=\frac{p^{L}}{(p^{2}+\beta^{2})^{M}}\sum_{n=0}^{n_{c}}
a^{(L)}_{in}P_{n}(x),\qquad
x=\frac{\beta^{2}-p^{2}}{\beta^{2}+p^{2}},
\label{eq:ffana}
\end{equation}
with $L$ the orbital angular momentum of the wave (the superscript is dropped where the wave is fixed), $P_{n}$ the Legendre polynomials and $n_{c}$ the order of the expansion.  Each form-factor component is then fitted with $n_{c}+1$ coefficients $a^{(L)}_{in}$ ($n_{c}=22$ for the nucleon--nucleon, $10$ for the $\alpha N$ and $12$ for the $\alpha\alpha$ separable potentials, Table~\ref{tab:ranks}); in a coupled channel each $\chi_{i}$ has one component of this form in each of the two waves, with the $L$ and $M$ of that wave. The integer $M$ controls the large-momentum behavior of the ansatz, $\chi_{i}(p)\sim p^{L-2M}$; it must be large enough for the propagator integrals $\mathcal D_{ij}(E)$ to converge and to match the decay of the exact half-shell form factor, and it is chosen per channel together with $\beta$ by the smallest $\max_{n}|a_{in}|$, which gives $M=1+\lfloor L/2\rfloor$, raised by one for the $^{3}F_{3}$ and $^{3}F_{4}$--$^{3}H_{4}$ separable potentials, whose momentum tails otherwise reach the quadrature nodes of fine three-body meshes. Table~\ref{tab:ranks} collects the resulting parameters. The coupling matrix is then \emph{rebuilt from the fitted factors} and symmetrized, so that the pair $(\chi,\bm{\Lambda}^{-1})$ entering the kernel is mutually consistent at the level of the fit residual.  Pauli-forbidden states, where present, are the bound states of the fitted representation itself. They are represented in the two-pole form
\begin{equation}
\phi_{fL}(p)\propto
\frac{p^{L}\sum_{n=0}^{n_{c}}A^{(f)}_{n}P_{n}(x)}{(p^{2}+\beta^{2})(p^{2}+\gamma_{f}^{2})},
\qquad
\gamma_{f}=\sqrt{2\mu|E_{f}|},
\label{eq:twopole}
\end{equation}
with $f$ labelling the state and $L$ its wave, $\gamma_{f}$ the bound-state momentum of the forbidden state at energy $E_{f}$ ($\mu$ the pair reduced mass), and $x$ the variable of Eq.~(\ref{eq:ffana}). For the channels with $M=1$ this is the exact bound state of the separable potential, the $A^{(f)}_{n}$ being the coefficients of the bound-state vector in the form factors~(\ref{eq:ffana}); for $M=2$ the single pole at $\beta$ is a fit of the exact form, which it reproduces accurately for the $\alpha\alpha$ $0d$ state and the $L=0$ states alike.

\begin{table}[htbp]
\caption{\label{tab:ranks}The separable potentials: rank $N_{r}$, Legendre order $n_{c}$ of Eq.~(\ref{eq:ffana}), range parameter $\beta$ and denominator power $M$ per component; $N_{f}$ is the number of Pauli-forbidden states.}
\begin{center}
\begin{tabular}{llcccc}
\toprule
potential & waves & $N_{r}$ & $n_{c}$ & $\beta$ & $M$ \\
\midrule
\multicolumn{6}{l}{CD~Bonn $NN$ (MeV units)} \\
$^{1}S_{0}$ ($np$, $pp$)           & $L=0$   & 5  & 22 & 500  & 1 \\
$^{1}S_{0}$ ($nn$)                 & $L=0$   & 7  & 22 & 900  & 1 \\
$^{3}S_{1}$--$^{3}D_{1}$ ($np$)    & $L=0,2$ & 9  & 22 & 1100 & 1, 2 \\
$^{3}P_{0}$, $^{3}P_{1}$ ($nn$, $np$, $pp$)& $L=1$  & 7  & 22 & 900  & 1 \\
$^{1}P_{1}$ ($np$)                 & $L=1$   & 6  & 22 & 900  & 1 \\
$^{1}D_{2}$ ($nn$, $np$, $pp$), $^{3}D_{2}$ ($np$) & $L=2$ & 6 & 22 & 900 & 2 \\
$^{3}F_{3}$ ($nn$, $np$, $pp$)            & $L=3$   & 6  & 22 & 900  & 3 \\
$^{1}F_{3}$ ($np$)                 & $L=3$   & 6  & 22 & 900  & 2 \\
$^{1}G_{4}$ ($nn$, $np$, $pp$), $^{3}G_{4}$ ($np$) & $L=4$ & 6 & 22 & 900 & 3 \\
$^{3}P_{2}$--$^{3}F_{2}$ ($nn$, $np$, $pp$) & $L=1,3$ & 14 & 22 & 900 & 1, 2 \\
$^{3}D_{3}$--$^{3}G_{3}$ ($np$)    & $L=2,4$ & 14 & 22 & 900  & 2, 3 \\
$^{3}F_{4}$--$^{3}H_{4}$ ($nn$, $np$, $pp$) & $L=3,5$ & 12 & 22 & 900 & 3, 4 \\
\multicolumn{6}{l}{KKNN $\alpha N$ (fm$^{-1}$ units; $N_{f}$ in parentheses)} \\
$^{2}S_{1/2}$ ($\alpha n$, $\alpha p$) & $L=0$   & 5 (1) & 10 & 2.0 & 1 \\
$^{2}P_{3/2}$, $^{2}P_{1/2}$               & $L=1$   & 3  & 10 & 1.5  & 1 \\
$^{2}D_{5/2}$, $^{2}D_{3/2}$               & $L=2$   & 3  & 10 & 1.5  & 2 \\
\multicolumn{6}{l}{BFW $\alpha\alpha$ (fm$^{-1}$ units)} \\
$L=0$                              & $L=0$   & 7 (2) & 12 & 3.0 & 1 \\
$L=2$                              & $L=2$   & 4 (1) & 12 & 3.0 & 2 \\
$L=4$                              & $L=4$   & 4  & 12 & 3.0  & 3 \\
\bottomrule
\end{tabular}
\end{center}
\end{table}

The ranks of Table~\ref{tab:ranks} are high beside the classical EST representations, which stayed at ranks of a few~\cite{Haidenbauer1984}; the placement of the supports, a bound state first and energies of both signs after it, follows that practice. The difference is the objective: the criteria of this paper demand the off-shell $t$~matrix over the whole domain the kernel samples, together with the thresholds, bound states and resonances, which fewer supports could not carry, and the modern low-rank factorizations of chiral interactions face the same trade-off between rank and accuracy~\cite{Tichai2019}. With separable input the Faddeev equations close on one-variable amplitudes of dimension $(\sum_{a}N_{r,a})\times N_{q}$, where $N_{r,a}$ are the channel ranks and $N_{q}$ the number of spectator-momentum mesh points, a small linear-algebra problem for present computers at every rank of Table~\ref{tab:ranks}. The high ranks appear only where the physics demands them: three suffices for the smooth $\alpha N$ waves, and the large ranks are confined to the tensor-coupled $NN$ channels, whose one-pion tensor structure is the hardest to represent at finite rank.

The complete set contains $30$ CD~Bonn separable potentials ($j\le4$: the eight $T=1$ waves in each of the $nn$, $np$, and $pp$ charge states and the six $T=0$ waves of the $np$ system), five KKNN separable potentials ($^{2}S_{1/2}$, $^{2}P_{3/2}$, $^{2}P_{1/2}$, $^{2}D_{5/2}$, $^{2}D_{3/2}$), and three BFW separable potentials ($L=0,2,4$). The coupled channels occur only in the $NN$ set, where the two components ($L$ and $L+2$) share the supports; the $T=1$ waves carry identical construction parameters in the three charge states, apart from $^{1}S_{0}$ $nn$ (Sec.~\ref{sec:1s0}). The separable potentials of Table~\ref{tab:ranks}, referred to below as the adopted set, with their Coulomb-dressed Pauli-forbidden states, are the complete two-body input of the three-body calculations of this work.\footnote{The adopted set is provided in machine-readable form as the ancillary file \texttt{est\_tables.dat} of the arXiv version of this paper, with a standalone script, \texttt{read\_back\_test.py}, that reads only that file and recomputes from it, under the conventions stated in the file header, for every block the phase shifts at three energies and the bound states, for the $^{1}S_{0}$ blocks the effective-range parameters and for the deuteron block the $D/S$ ratio and the $D$-state probability, and compares five blocks in detail with the values quoted in their comments. The file holds the CD~Bonn blocks of every $j\le4$ wave in all three charge states, the KKNN $\alpha n$ blocks together with the $\alpha p$ blocks of the particle basis (the $\alpha n$ fit at the $\alpha p$ reduced mass) and the BFW blocks, all in the sign convention of the three-body kernel (Sec.~\ref{sec:dsign}), the forbidden states as the reduced form factors $\phi_{fL}/p^{L}$, the factor $p^{L}$ being restored in the kernel. The support energies of each representation and its per-channel acceptance numbers for the criteria C1--C6 are listed with the corresponding block.}

\section{The nucleon--nucleon interaction}
\label{sec:nn}

\subsection{The CD~Bonn potential and its partial waves}
\label{sec:np}

The CD~Bonn potential was constructed from its published parameters~\cite{Machleidt2001} and reproduces the published phase shifts, mixing angles, deuteron properties and effective-range parameters to the rounding of the published tables; Table~\ref{tab:parent} lists the deuteron properties and effective-range parameters of the potential as constructed here, the reference for the comparisons below. The known exception is the deuteron quadrupole moment, which CD~Bonn itself underpredicts; the difference is attributed to meson-exchange currents and relativistic corrections outside a potential-model calculation.

\begin{table}[htbp]
\caption{\label{tab:parent}Verification of the CD~Bonn potential as constructed here: deuteron properties ($B_{d}$ the binding energy, $P_{D}$ the $D$-state probability, $\eta$ the asymptotic $D/S$ ratio of the wave function) and effective-range parameters; $a_{pp}^{N}$, $r_{pp}^{N}$ refer to the nuclear part of the $pp$ interaction.}
\begin{center}
\begin{tabular}{lc}
\toprule
quantity & value \\
\midrule
$B_{d}$ (MeV) & $2.224574$ \\
$P_{D}$ (\%) & $4.854$ \\
$\eta$ & $0.0256$ \\
$a_{nn}$, $r_{nn}$ (fm) & $-18.9680$, $2.8182$ \\
$a_{np}$, $r_{np}$ (fm) & $-23.7380$, $2.6701$ \\
$a_{pp}^{N}$, $r_{pp}^{N}$ (fm) & $-17.4601$, $2.8441$ \\
\bottomrule
\end{tabular}
\end{center}
\end{table}  
The $np$ system contains more of the potential set than the other two charge states. Table~\ref{tab:npwaves} sorts its fourteen $j\le4$ partial waves by isospin. The eight $T=1$ waves are tabulated separately for $nn$, $np$, and $pp$, so the charge dependence of CD~Bonn is carried by the separable potentials themselves, and the fitted $^{1}S_{0}$ separable potentials reproduce the scattering lengths of the original potential in all three charge states almost exactly (Sec.~\ref{sec:1s0}). The six $T=0$ waves exist for $np$ only. Among them the coupled $^{3}S_{1}$--$^{3}D_{1}$ channel carries the deuteron, and it is the channel on which the most sensitive criteria of this paper act: the $\eta$ criterion (Sec.~\ref{sec:eta}) and the three-body sign condition (Sec.~\ref{sec:dsign}) are both statements about the $np$ triplet separable potential.

\begin{table}[htbp]
\caption{\label{tab:npwaves}The fourteen $j\le4$ partial waves of the $np$ system by isospin. The $T=1$ waves exist in all three charge states and carry the charge dependence of CD~Bonn; the $T=0$ waves exist for $np$ only. Coupled channels are joined by a dash.}
\begin{center}
\begin{tabular}{cp{7.2cm}cc}
\toprule
$T$ & partial waves & charge states & number \\
\midrule
$1$ & $^{1}S_{0}$, $^{3}P_{0}$, $^{3}P_{1}$, $^{1}D_{2}$, $^{3}F_{3}$, $^{1}G_{4}$, $^{3}P_{2}$--$^{3}F_{2}$, $^{3}F_{4}$--$^{3}H_{4}$ & $nn$, $np$, $pp$ & 8 \\
$0$ & $^{3}S_{1}$--$^{3}D_{1}$, $^{1}P_{1}$, $^{3}D_{2}$, $^{1}F_{3}$, $^{3}G_{4}$, $^{3}D_{3}$--$^{3}G_{3}$ & $np$ & 6 \\
\bottomrule
\end{tabular}
\end{center}
\end{table}  
The $np$ separable potentials enter every system treated in this work in which a proton is present ($^{3}$H, $^{3}$He, $^{6}$Li). Together with the $nn$ separable potentials they are the complete input of the neutron--deuteron calculations of Sec.~\ref{sec:threebody}.

\subsection{Phase shifts of the separable representations}
\label{sec:nnphases}

Criterion C2 for the nucleon--nucleon set is the comparison of the separable representations with the original potential between the support points; for the $pp$ potentials the phases are the nuclear ones, without the Coulomb interaction, on both sides. Figure~\ref{fig:nnphases} shows it for the four waves which the later subsections examine in detail, the $^{1}S_{0}$ $np$ and $pp$ potentials (Sec.~\ref{sec:1s0}) and the $^{3}S_{1}$ and $^{3}D_{1}$ phases of the coupled channel (Sec.~\ref{sec:eta}).

\begin{figure}[htbp]
\begin{center}
\begin{tikzpicture}
\begin{groupplot}[group style={group size=2 by 2, horizontal sep=1.5cm, vertical sep=1.6cm},
  width=0.48\textwidth, height=5.0cm, xmode=log, log basis x=10,
  xlabel={$T_{\rm lab}$ (MeV)}, xtick={1,10,100,300}, xticklabels={1,10,100,300},
  tick label style={font=\scriptsize}, label style={font=\scriptsize},
  title style={font=\scriptsize}, every mark/.append style={scale=0.9},
  legend style={font=\scriptsize, draw=none, fill=none, inner sep=1pt, row sep=-3pt}, legend cell align=left]
\nextgroupplot[title={$^{1}S_{0}$ ($np$)}, ylabel={$\delta$ (deg)}, legend pos=south west]
\addplot[black, forget plot] coordinates {(1,62.090) (2,64.490) (5,63.670) (10,60.010) (25,50.930) (50,40.450) (100,26.380) (150,16.320) (200,8.310) (300,-4.250)};
\addplot[black, only marks, mark=*] coordinates {(1,62.090) (2,64.490) (5,63.670) (10,60.000) (25,50.930) (50,40.430) (100,26.360) (150,16.270) (200,8.090) (300,-4.370)}; \addlegendentry{separable}
\nextgroupplot[title={$^{1}S_{0}$ ($pp$)}]
\addplot[black, forget plot] coordinates {(1,55.900) (2,59.560) (5,60.070) (10,57.100) (25,48.580) (50,38.270) (100,24.130) (150,13.930) (200,5.780) (300,-7.040)};
\addplot[black, only marks, mark=*] coordinates {(1,55.900) (2,59.560) (5,60.070) (10,57.100) (25,48.580) (50,38.230) (100,24.110) (150,13.880) (200,5.540) (300,-7.160)};
\nextgroupplot[title={$^{3}S_{1}$}, ylabel={$\delta$ (deg)}]
\addplot[black, forget plot] coordinates {(1,147.750) (2,136.460) (5,118.180) (10,102.620) (25,80.630) (50,62.730) (100,43.060) (150,30.470) (200,20.950) (300,6.650)};
\addplot[black, only marks, mark=*] coordinates {(1,147.700) (2,136.400) (5,118.080) (10,102.520) (25,80.590) (50,62.700) (100,42.770) (150,30.150) (200,20.640) (300,6.650)};
\nextgroupplot[title={$^{3}D_{1}$}]
\addplot[black, forget plot] coordinates {(1,-0.005) (2,-0.026) (5,-0.184) (10,-0.678) (25,-2.803) (50,-6.440) (100,-12.255) (150,-16.495) (200,-19.679) (300,-24.027)};
\addplot[black, only marks, mark=*] coordinates {(1,-0.002) (2,-0.012) (5,-0.090) (10,-0.380) (25,-2.080) (50,-5.914) (100,-12.236) (150,-15.701) (200,-18.631) (300,-23.994)};
\end{groupplot}
\end{tikzpicture}
\end{center}
\caption{Phase shifts of the CD~Bonn potential (lines) and of the separable representations (points), criterion C2, against the laboratory energy: the $^{1}S_{0}$ $np$ and $pp$ potentials ($pp$ without the Coulomb interaction) and the $^{3}S_{1}$ and $^{3}D_{1}$ phases of the coupled channel.}
\label{fig:nnphases}
\end{figure}

\begin{figure}[p]
\begin{center}
\begin{tikzpicture}
\begin{groupplot}[group style={group size=3 by 4, horizontal sep=1.3cm, vertical sep=2.0cm},
  width=0.36\textwidth, height=4.2cm, xmode=log, log basis x=10,
  xlabel={$T_{\rm lab}$ (MeV)}, xtick={1,10,100,300}, xticklabels={1,10,100,300},
  tick label style={font=\scriptsize}, label style={font=\scriptsize},
  title style={font=\scriptsize}, every mark/.append style={scale=0.7},
  legend style={font=\scriptsize, draw=none, fill=none, inner sep=1pt, row sep=-3pt}, legend cell align=left]
\nextgroupplot[title={$^{3}P_{0}$}, ylabel={$\delta$ (deg)}, legend pos=north west]
\addplot[black, forget plot] coordinates {(1,0.177) (2,0.474) (5,1.608) (10,3.621) (25,8.100) (50,10.738) (100,8.572) (150,3.719) (200,-1.547) (300,-11.537)};
\addplot[black, only marks, mark=*] coordinates {(1,0.176) (2,0.474) (5,1.612) (10,3.625) (25,8.100) (50,10.700) (100,8.526) (150,3.530) (200,-1.626) (300,-11.539)}; \addlegendentry{$np$}
\addplot[black, forget plot] coordinates {(1,0.207) (2,0.551) (5,1.841) (10,4.078) (25,8.901) (50,11.652) (100,9.425) (150,4.504) (200,-0.807) (300,-10.848)};
\addplot[black, only marks, mark=o, mark options={scale=1.7}] coordinates {(1,0.206) (2,0.550) (5,1.848) (10,4.085) (25,8.902) (50,11.600) (100,9.364) (150,4.256) (200,-0.906) (300,-10.850)}; \addlegendentry{$pp$}
\addplot[black, forget plot] coordinates {(1,0.208) (2,0.553) (5,1.849) (10,4.096) (25,8.941) (50,11.707) (100,9.486) (150,4.562) (200,-0.753) (300,-10.800)};
\addplot[black, only marks, mark=triangle, mark options={scale=1.5}] coordinates {(1,0.206) (2,0.552) (5,1.856) (10,4.103) (25,8.942) (50,11.655) (100,9.424) (150,4.315) (200,-0.849) (300,-10.802)}; \addlegendentry{$nn$}
\nextgroupplot[title={$^{3}P_{1}$}]
\addplot[black, forget plot] coordinates {(1,-0.107) (2,-0.282) (5,-0.925) (10,-2.036) (25,-4.813) (50,-8.181) (100,-13.234) (150,-17.507) (200,-21.382) (300,-28.273)};
\addplot[black, only marks, mark=*] coordinates {(1,-0.107) (2,-0.282) (5,-0.928) (10,-2.039) (25,-4.815) (50,-8.161) (100,-13.197) (150,-17.179) (200,-21.046) (300,-28.272)};
\addplot[black, forget plot] coordinates {(1,-0.124) (2,-0.323) (5,-1.037) (10,-2.234) (25,-5.129) (50,-8.553) (100,-13.633) (150,-17.912) (200,-21.789) (300,-28.683)};
\addplot[black, only marks, mark=o, mark options={scale=1.7}] coordinates {(1,-0.123) (2,-0.322) (5,-1.042) (10,-2.239) (25,-5.131) (50,-8.529) (100,-13.588) (150,-17.533) (200,-21.411) (300,-28.681)};
\addplot[black, forget plot] coordinates {(1,-0.124) (2,-0.323) (5,-1.038) (10,-2.235) (25,-5.127) (50,-8.542) (100,-13.604) (150,-17.871) (200,-21.741) (300,-28.627)};
\addplot[black, only marks, mark=triangle, mark options={scale=1.5}] coordinates {(1,-0.123) (2,-0.323) (5,-1.043) (10,-2.241) (25,-5.129) (50,-8.517) (100,-13.558) (150,-17.483) (200,-21.350) (300,-28.626)};
\nextgroupplot[title={$^{1}D_{2}$}]
\addplot[black, forget plot] coordinates {(1,0.001) (2,0.006) (5,0.042) (10,0.158) (25,0.687) (50,1.726) (100,3.865) (150,5.816) (200,7.455) (300,9.759)};
\addplot[black, only marks, mark=*] coordinates {(1,0.001) (2,0.004) (5,0.036) (10,0.150) (25,0.684) (50,1.686) (100,3.824) (150,5.550) (200,7.252) (300,9.758)};
\addplot[black, forget plot] coordinates {(1,0.001) (2,0.007) (5,0.050) (10,0.182) (25,0.734) (50,1.764) (100,3.846) (150,5.747) (200,7.348) (300,9.610)};
\addplot[black, only marks, mark=o, mark options={scale=1.7}] coordinates {(1,0.001) (2,0.005) (5,0.041) (10,0.169) (25,0.730) (50,1.713) (100,3.797) (150,5.434) (200,7.113) (300,9.609)};
\addplot[black, forget plot] coordinates {(1,0.001) (2,0.007) (5,0.051) (10,0.183) (25,0.738) (50,1.775) (100,3.877) (150,5.801) (200,7.423) (300,9.722)};
\addplot[black, only marks, mark=triangle, mark options={scale=1.5}] coordinates {(1,0.001) (2,0.005) (5,0.041) (10,0.170) (25,0.734) (50,1.724) (100,3.828) (150,5.487) (200,7.187) (300,9.721)};
\nextgroupplot[title={$^{3}P_{2}$}, ylabel={$\delta$ (deg)}]
\addplot[black, forget plot] coordinates {(1,0.022) (2,0.064) (5,0.257) (10,0.722) (25,2.596) (50,5.934) (100,11.014) (150,13.984) (200,15.660) (300,17.080)};
\addplot[black, only marks, mark=*] coordinates {(1,0.028) (2,0.077) (5,0.282) (10,0.748) (25,2.604) (50,6.223) (100,11.350) (150,14.690) (200,15.955) (300,17.584)};
\addplot[black, forget plot] coordinates {(1,0.023) (2,0.066) (5,0.266) (10,0.751) (25,2.687) (50,6.096) (100,11.226) (150,14.198) (200,15.863) (300,17.255)};
\addplot[black, only marks, mark=o, mark options={scale=1.7}] coordinates {(1,0.030) (2,0.081) (5,0.296) (10,0.783) (25,2.697) (50,6.451) (100,11.643) (150,15.019) (200,16.197) (300,17.811)};
\addplot[black, forget plot] coordinates {(1,0.023) (2,0.066) (5,0.269) (10,0.758) (25,2.711) (50,6.150) (100,11.326) (150,14.325) (200,16.006) (300,17.415)};
\addplot[black, only marks, mark=triangle, mark options={scale=1.5}] coordinates {(1,0.031) (2,0.083) (5,0.306) (10,0.800) (25,2.724) (50,6.591) (100,11.847) (150,15.455) (200,16.371) (300,18.105)};
\nextgroupplot[title={$^{3}F_{2}$}]
\addplot[black, forget plot] coordinates {(1,0.000) (2,0.000) (5,0.002) (10,0.011) (25,0.089) (50,0.297) (100,0.720) (150,1.031) (200,1.188) (300,0.981)};
\addplot[black, only marks, mark=*] coordinates {(1,0.000) (2,0.000) (5,0.001) (10,0.004) (25,0.057) (50,0.275) (100,0.727) (150,1.176) (200,1.221) (300,0.969)};
\addplot[black, forget plot] coordinates {(1,0.000) (2,0.000) (5,0.002) (10,0.015) (25,0.109) (50,0.340) (100,0.786) (150,1.107) (200,1.270) (300,1.071)};
\addplot[black, only marks, mark=o, mark options={scale=1.7}] coordinates {(1,0.000) (2,0.000) (5,0.001) (10,0.005) (25,0.060) (50,0.299) (100,0.791) (150,1.308) (200,1.307) (300,1.051)};
\addplot[black, forget plot] coordinates {(1,0.000) (2,0.000) (5,0.002) (10,0.015) (25,0.109) (50,0.341) (100,0.790) (150,1.114) (200,1.280) (300,1.086)};
\addplot[black, only marks, mark=triangle, mark options={scale=1.5}] coordinates {(1,0.000) (2,0.000) (5,0.001) (10,0.004) (25,0.054) (50,0.290) (100,0.790) (150,1.297) (200,1.306) (300,1.061)};
\nextgroupplot[title={$\bar\varepsilon_{2}$}]
\addplot[black, forget plot] coordinates {(1,-0.001) (2,-0.007) (5,-0.049) (10,-0.184) (25,-0.756) (50,-1.639) (100,-2.628) (150,-2.899) (200,-2.795) (300,-2.128)};
\addplot[black, only marks, mark=*] coordinates {(1,-0.001) (2,-0.004) (5,-0.033) (10,-0.149) (25,-0.792) (50,-1.707) (100,-2.687) (150,-3.006) (200,-2.726) (300,-2.277)};
\addplot[black, forget plot] coordinates {(1,-0.002) (2,-0.009) (5,-0.060) (10,-0.217) (25,-0.842) (50,-1.754) (100,-2.728) (150,-2.973) (200,-2.850) (300,-2.162)};
\addplot[black, only marks, mark=o, mark options={scale=1.7}] coordinates {(1,-0.001) (2,-0.004) (5,-0.037) (10,-0.169) (25,-0.893) (50,-1.832) (100,-2.796) (150,-3.066) (200,-2.757) (300,-2.334)};
\addplot[black, forget plot] coordinates {(1,-0.002) (2,-0.009) (5,-0.060) (10,-0.218) (25,-0.845) (50,-1.758) (100,-2.732) (150,-2.973) (200,-2.846) (300,-2.149)};
\addplot[black, only marks, mark=triangle, mark options={scale=1.5}] coordinates {(1,-0.001) (2,-0.004) (5,-0.037) (10,-0.170) (25,-0.896) (50,-1.822) (100,-2.793) (150,-2.959) (200,-2.744) (300,-2.341)};
\nextgroupplot[title={$^{3}F_{3}$}, ylabel={$\delta$ (deg)}]
\addplot[black, forget plot] coordinates {(1,-0.000) (2,-0.000) (5,-0.004) (10,-0.027) (25,-0.199) (50,-0.623) (100,-1.417) (150,-1.979) (200,-2.329) (300,-2.570)};
\addplot[black, only marks, mark=*] coordinates {(1,-0.000) (2,-0.000) (5,-0.001) (10,-0.013) (25,-0.163) (50,-0.622) (100,-1.334) (150,-1.974) (200,-2.295) (300,-2.568)};
\addplot[black, forget plot] coordinates {(1,-0.000) (2,-0.000) (5,-0.006) (10,-0.035) (25,-0.241) (50,-0.710) (100,-1.551) (150,-2.134) (200,-2.496) (300,-2.755)};
\addplot[black, only marks, mark=o, mark options={scale=1.7}] coordinates {(1,-0.000) (2,-0.000) (5,-0.002) (10,-0.015) (25,-0.190) (50,-0.710) (100,-1.444) (150,-2.127) (200,-2.453) (300,-2.753)};
\addplot[black, forget plot] coordinates {(1,-0.000) (2,-0.000) (5,-0.006) (10,-0.035) (25,-0.242) (50,-0.712) (100,-1.553) (150,-2.134) (200,-2.492) (300,-2.738)};
\addplot[black, only marks, mark=triangle, mark options={scale=1.5}] coordinates {(1,-0.000) (2,-0.000) (5,-0.002) (10,-0.015) (25,-0.191) (50,-0.712) (100,-1.446) (150,-2.126) (200,-2.448) (300,-2.736)};
\nextgroupplot[title={$^{1}G_{4}$}]
\addplot[black, forget plot] coordinates {(1,0.000) (2,0.000) (5,0.000) (10,0.003) (25,0.032) (50,0.134) (100,0.391) (150,0.669) (200,0.961) (300,1.562)};
\addplot[black, only marks, mark=*] coordinates {(1,0.000) (2,0.000) (5,0.000) (10,0.001) (25,0.024) (50,0.134) (100,0.352) (150,0.667) (200,0.943) (300,1.561)};
\addplot[black, forget plot] coordinates {(1,0.000) (2,0.000) (5,0.000) (10,0.004) (25,0.041) (50,0.155) (100,0.421) (150,0.695) (200,0.979) (300,1.560)};
\addplot[black, only marks, mark=o, mark options={scale=1.7}] coordinates {(1,0.000) (2,0.000) (5,0.000) (10,0.001) (25,0.028) (50,0.156) (100,0.372) (150,0.691) (200,0.956) (300,1.559)};
\addplot[black, forget plot] coordinates {(1,0.000) (2,0.000) (5,0.000) (10,0.004) (25,0.042) (50,0.156) (100,0.423) (150,0.699) (200,0.985) (300,1.572)};
\addplot[black, only marks, mark=triangle, mark options={scale=1.5}] coordinates {(1,0.000) (2,0.000) (5,0.000) (10,0.001) (25,0.028) (50,0.156) (100,0.373) (150,0.695) (200,0.962) (300,1.571)};
\nextgroupplot[title={$^{3}F_{4}$}]
\addplot[black, forget plot] coordinates {(1,0.000) (2,0.000) (5,0.000) (10,0.001) (25,0.018) (50,0.106) (100,0.478) (150,1.007) (200,1.586) (300,2.645)};
\addplot[black, only marks, mark=*] coordinates {(1,0.000) (2,0.000) (5,0.000) (10,0.001) (25,0.016) (50,0.103) (100,0.476) (150,1.006) (200,1.584) (300,2.647)};
\addplot[black, forget plot] coordinates {(1,0.000) (2,0.000) (5,0.000) (10,0.002) (25,0.022) (50,0.121) (100,0.513) (150,1.057) (200,1.646) (300,2.715)};
\addplot[black, only marks, mark=o, mark options={scale=1.7}] coordinates {(1,0.000) (2,0.000) (5,0.000) (10,0.001) (25,0.020) (50,0.111) (100,0.509) (150,1.056) (200,1.643) (300,2.717)};
\addplot[black, forget plot] coordinates {(1,0.000) (2,0.000) (5,0.000) (10,0.002) (25,0.022) (50,0.122) (100,0.516) (150,1.063) (200,1.656) (300,2.731)};
\addplot[black, only marks, mark=triangle, mark options={scale=1.5}] coordinates {(1,0.000) (2,0.000) (5,0.000) (10,0.001) (25,0.020) (50,0.112) (100,0.512) (150,1.062) (200,1.652) (300,2.733)};
\nextgroupplot[title={$^{3}H_{4}$}, ylabel={$\delta$ (deg)}]
\addplot[black, forget plot] coordinates {(1,0.000) (2,0.000) (5,0.000) (10,0.000) (25,0.003) (50,0.021) (100,0.094) (150,0.196) (200,0.311) (300,0.541)};
\addplot[black, only marks, mark=*] coordinates {(1,0.000) (2,0.000) (5,0.000) (10,0.000) (25,0.010) (50,-0.068) (100,0.092) (150,0.108) (200,0.256) (300,0.520)};
\addplot[black, forget plot] coordinates {(1,0.000) (2,0.000) (5,0.000) (10,0.000) (25,0.004) (50,0.027) (100,0.112) (150,0.223) (200,0.345) (300,0.582)};
\addplot[black, only marks, mark=o, mark options={scale=1.7}] coordinates {(1,0.000) (2,0.000) (5,0.000) (10,0.000) (25,0.006) (50,-0.030) (100,0.118) (150,0.181) (200,0.301) (300,0.558)};
\addplot[black, forget plot] coordinates {(1,0.000) (2,0.000) (5,0.000) (10,0.000) (25,0.004) (50,0.027) (100,0.112) (150,0.224) (200,0.346) (300,0.586)};
\addplot[black, only marks, mark=triangle, mark options={scale=1.5}] coordinates {(1,0.000) (2,0.000) (5,0.000) (10,0.000) (25,0.006) (50,-0.030) (100,0.119) (150,0.182) (200,0.303) (300,0.561)};
\nextgroupplot[title={$\bar\varepsilon_{4}$}]
\addplot[black, forget plot] coordinates {(1,-0.000) (2,-0.000) (5,-0.000) (10,-0.003) (25,-0.039) (50,-0.167) (100,-0.490) (150,-0.791) (200,-1.045) (300,-1.416)};
\addplot[black, only marks, mark=*] coordinates {(1,-0.000) (2,-0.000) (5,-0.000) (10,-0.000) (25,-0.015) (50,-0.152) (100,-0.474) (150,-0.775) (200,-1.091) (300,-1.487)};
\addplot[black, forget plot] coordinates {(1,-0.000) (2,-0.000) (5,-0.000) (10,-0.004) (25,-0.051) (50,-0.199) (100,-0.545) (150,-0.854) (200,-1.110) (300,-1.476)};
\addplot[black, only marks, mark=o, mark options={scale=1.7}] coordinates {(1,-0.000) (2,-0.000) (5,-0.000) (10,-0.000) (25,-0.015) (50,-0.181) (100,-0.526) (150,-0.835) (200,-1.166) (300,-1.560)};
\addplot[black, forget plot] coordinates {(1,-0.000) (2,-0.000) (5,-0.000) (10,-0.004) (25,-0.051) (50,-0.199) (100,-0.546) (150,-0.857) (200,-1.113) (300,-1.480)};
\addplot[black, only marks, mark=triangle, mark options={scale=1.5}] coordinates {(1,-0.000) (2,-0.000) (5,-0.000) (10,-0.000) (25,-0.015) (50,-0.182) (100,-0.528) (150,-0.837) (200,-1.168) (300,-1.566)};
\end{groupplot}
\end{tikzpicture}
\end{center}
\caption{Phase shifts and mixing parameters of the $T=1$ partial waves with $j\le4$: the CD~Bonn potential (lines, $np$) and the separable representations (points: circles $np$, open circles $pp$, triangles $nn$), criterion C2, against the laboratory energy; the $pp$ phases are the nuclear ones, without the Coulomb interaction. The charge dependence of the original potential outside $^{1}S_{0}$ is below the width of the lines.}
\label{fig:nnT1}
\end{figure}

\begin{figure}[htbp]
\begin{center}
\begin{tikzpicture}
\begin{groupplot}[group style={group size=3 by 3, horizontal sep=1.3cm, vertical sep=2.0cm},
  width=0.36\textwidth, height=4.2cm, xmode=log, log basis x=10,
  xlabel={$T_{\rm lab}$ (MeV)}, xtick={1,10,100,300}, xticklabels={1,10,100,300},
  tick label style={font=\scriptsize}, label style={font=\scriptsize},
  title style={font=\scriptsize}, every mark/.append style={scale=0.7},
  legend style={font=\scriptsize, draw=none, fill=none, inner sep=1pt, row sep=-3pt}, legend cell align=left]
\nextgroupplot[title={$^{1}P_{1}$}, ylabel={$\delta$ (deg)}]
\addplot[black, forget plot] coordinates {(1,-0.188) (2,-0.483) (5,-1.490) (10,-3.050) (25,-6.348) (50,-9.726) (100,-14.425) (150,-18.326) (200,-21.775) (300,-27.571)};
\addplot[black, only marks, mark=*] coordinates {(1,-0.185) (2,-0.481) (5,-1.500) (10,-3.063) (25,-6.348) (50,-9.675) (100,-14.321) (150,-17.508) (200,-21.022) (300,-27.566)};
\nextgroupplot[title={$^{3}D_{2}$}]
\addplot[black, forget plot] coordinates {(1,0.006) (2,0.030) (5,0.223) (10,0.850) (25,3.722) (50,8.971) (100,17.222) (150,22.089) (200,24.510) (300,25.214)};
\addplot[black, only marks, mark=*] coordinates {(1,0.004) (2,0.023) (5,0.187) (10,0.799) (25,3.702) (50,8.705) (100,16.948) (150,20.315) (200,23.168) (300,25.208)};
\nextgroupplot[title={$^{1}F_{3}$}]
\addplot[black, forget plot] coordinates {(1,-0.000) (2,-0.001) (5,-0.011) (10,-0.066) (25,-0.415) (50,-1.106) (100,-2.152) (150,-2.871) (200,-3.478) (300,-4.734)};
\addplot[black, only marks, mark=*] coordinates {(1,-0.000) (2,-0.000) (5,-0.003) (10,-0.025) (25,-0.317) (50,-1.106) (100,-1.963) (150,-2.858) (200,-3.422) (300,-4.731)};
\nextgroupplot[title={$^{3}D_{3}$}, ylabel={$\delta$ (deg)}]
\addplot[black, forget plot] coordinates {(1,0.000) (2,0.000) (5,0.002) (10,0.007) (25,0.052) (50,0.333) (100,1.452) (150,2.704) (200,3.696) (300,4.538)};
\addplot[black, only marks, mark=*] coordinates {(1,0.000) (2,0.003) (5,-0.001) (10,-0.002) (25,0.048) (50,0.362) (100,1.502) (150,3.040) (200,3.788) (300,4.582)};
\nextgroupplot[title={$^{3}G_{3}$}]
\addplot[black, forget plot] coordinates {(1,-0.000) (2,-0.000) (5,-0.000) (10,-0.004) (25,-0.053) (50,-0.259) (100,-0.939) (150,-1.762) (200,-2.599) (300,-4.094)};
\addplot[black, only marks, mark=*] coordinates {(1,0.000) (2,0.004) (5,-0.004) (10,-0.017) (25,0.035) (50,0.206) (100,-0.763) (150,-0.507) (200,-2.122) (300,-3.913)};
\nextgroupplot[title={$\bar\varepsilon_{3}$}]
\addplot[black, forget plot] coordinates {(1,0.000) (2,0.001) (5,0.013) (10,0.081) (25,0.551) (50,1.607) (100,3.486) (150,4.834) (200,5.763) (300,6.835)};
\addplot[black, only marks, mark=*] coordinates {(1,0.000) (2,0.004) (5,-0.005) (10,-0.051) (25,0.601) (50,1.829) (100,3.590) (150,5.618) (200,6.273) (300,7.116)};
\nextgroupplot[title={$^{3}G_{4}$}, ylabel={$\delta$ (deg)}]
\addplot[black, forget plot] coordinates {(1,0.000) (2,0.000) (5,0.001) (10,0.014) (25,0.170) (50,0.721) (100,2.173) (150,3.640) (200,4.986) (300,7.206)};
\addplot[black, only marks, mark=*] coordinates {(1,0.000) (2,0.000) (5,0.000) (10,0.004) (25,0.120) (50,0.721) (100,1.977) (150,3.625) (200,4.885) (300,7.202)};
\end{groupplot}
\end{tikzpicture}
\end{center}
\caption{Phase shifts and the mixing parameter $\bar\varepsilon_{3}$ of the $T=0$ partial waves with $j\le4$ other than $^{3}S_{1}$--$^{3}D_{1}$ (Fig.~\ref{fig:nnphases}) and $\bar\varepsilon_{1}$ (Sec.~\ref{sec:eta}): the CD~Bonn potential (lines) and the separable representations (points), criterion C2, against the laboratory energy.}
\label{fig:nnT0}
\end{figure}

Figures~\ref{fig:nnT1} and~\ref{fig:nnT0} compare the separable representations with the original potential in the remaining twenty-six waves and mixing parameters, sorted by isospin as in Table~\ref{tab:npwaves}. Every phase and mixing parameter of every charge state is reproduced closely below $100$~MeV, except $\bar\varepsilon_{1}$ at $10$--$25$~MeV and $^{3}D_{1}$ at $25$~MeV; above $100$~MeV the deviations reach about two degrees, in the partner waves and mixing parameters of the coupled channels and in $^{3}P_{2}$, $^{1}P_{1}$ and $^{3}D_{2}$ at $150$--$200$~MeV. The partner waves of the coupled channels are the least constrained by the supports, and they are not negligible in the three-body problem: a $5^{\circ}$ error in a partner wave whose phase is itself below $1.3^{\circ}$ is worth $4$~keV in the $j\le4$ triton (Table~\ref{tab:sens}).

\subsection{The coupled channel: $\eta$ as the most sensitive test}
\label{sec:eta}

Table~\ref{tab:eta} presents the central result for the nucleon--nucleon interaction. The rank-$9$ $^{3}S_{1}$--$^{3}D_{1}$ representation takes the deuteron vertex function of the original potential as its first support, with the remaining supports at negative energies; it reproduces $B_{d}$ (Table~\ref{tab:eta}) and the low-energy phases accurately. The asymptotic ratio is recomputed from the fitted separable potential in three steps: the zero of $\det[\bm{\Lambda}^{-1}-\bm{\mathcal D}(E)]$, the deuteron pole, is located, the null vector $c$ is extracted, and the vertex functions
\begin{equation}
G_{L}(p)=\sum_{i}c_{i}\,\chi_{i}^{(L)}(p)
\label{eq:vertex}
\end{equation}
with $\chi_{i}^{(L)}$ the $L=0$ and $L=2$ components~(\ref{eq:ffana}) of the form factors, are continued to the pole, $p^{2}=-\gamma^{2}$ with $\gamma=\sqrt{2\mu B_{d}}$ the deuteron wave number. The bound-state wave function of a separable potential is $\psi_{L}(p)=G_{L}(p)/(E-p^{2}/2\mu)$, so its asymptotic normalization constants are the residues of the vertex functions at the pole, and
\begin{equation}
\eta=\frac{\gamma^{2}\,\hat G_{2}(-\gamma^{2})}{\hat G_{0}(-\gamma^{2})},
\qquad G_{L}(p)=p^{L}\,\hat G_{L}(p^{2}),
\label{eq:etadef}
\end{equation}
the factor $p^{L}$ being continued as $\gamma^{L}$, which fixes the phase convention in which $A_{S}$ and $A_{D}$ are both positive. The procedure is exact and requires no coordinate-space fitting.  The result, together with the off-shell measure (\ref{eq:eps}), is given in Table~\ref{tab:eta}: $\eta$ and $P_{D}$ agree with the original potential to the quoted digits, and the off-shell measure lies between $1.2\%$ at $-5$~MeV and $6.7\%$ at $-200$~MeV.

\begin{table}[htbp]
\caption{\label{tab:eta}The $^{3}S_{1}$--$^{3}D_{1}$ channel: the rank-9 separable representation against the original potential.}
\begin{center}
\begin{tabular}{lcc}
\toprule
quantity & separable & original \\
\midrule
$B_{d}$ (MeV)             & $2.223256$ & $2.224574$ \\
$\eta$                    & $0.02556$  & $0.0256$   \\
$P_{D}$ (\%)              & $4.854$    & $4.854$    \\
$\Xi(-5\,\mathrm{MeV})$   & $1.2\%$ & --- \\
$\Xi(-200\,\mathrm{MeV})$ & $6.7\%$ & --- \\
\bottomrule
\end{tabular}
\end{center}
\end{table}
The sensitivity of this channel to the placement of the supports is the reason for the choice: the binding energy, the $D$-state probability and the phase shifts at the support energies do not fix the residues at the pole, and a rank-9 representation which reproduces $B_{d}$ and the low-energy phases equally well can still miss $\eta$ by $3.4\%$, with a seven- to tenfold larger off-shell measure at the same rank, which is worth about $90$~keV in the triton and in $^{6}$Li (Table~\ref{tab:sens}). The representation adopted here, with the deuteron vertex function itself as first support and the remaining supports at negative energies, has $\eta$ and $P_{D}$ at the values of the original potential together with the smaller off-shell measure (Fig.~\ref{fig:c1nn}). Its weakest on-shell aspect is the mixing parameter: $\bar\varepsilon_{1}$ deviates from the original potential by up to about a degree at $10$--$25$~MeV and about two degrees near $150$--$200$~MeV, a support-placement effect, the representation having no support on the positive axis; the coupled $^{3}S_{1}$--$^{3}D_{1}$ channel was the one needing the most care in separable representations of realistic potentials from the beginning, the Paris representation requiring rank $4$ for it~\cite{Haidenbauer1984}. The mixing parameter of the representation at three energies, in the Stapp convention, is also quoted with its block in the adopted set as the check of the stored sign of the $D$ component. The bound-state-relevant mixing is fixed instead by the exact $\eta$ at the deuteron pole, and an $\varepsilon_{1}$ constraint at high energy, with the deuteron constraints unchanged, moves the triton and $^{6}$Li by a keV or two only (Table~\ref{tab:sens}).

\subsection{The $^{1}S_{0}$ channels: effective-range parameters under negative-energy optimization}
\label{sec:1s0}

The rank-$5$ $^{1}S_{0}$ $np$ and $pp$ representations, built on positive-energy supports, satisfy C3 with a large margin but deviate from the original $t$~matrix by $28$--$44\%$ over the kernel energies (Fig.~\ref{fig:c1nn}), against $0.7\%$ for the rank-$7$ $nn$ potential, which is built on negative-energy supports together with one support at $E=0.15$~MeV that fixes the threshold (Table~\ref{tab:s0thr}). The low-energy support is necessary: negative-energy supports alone optimize the amplitude at the kernel energies at the expense of the threshold region, and neither the on-shell nor the off-shell diagnostics signal it; without the low-energy support the scattering length of this channel can be off by $10\%$ while the off-shell measure stays below one percent, which is worth $15$~keV in $^{6}$He (Table~\ref{tab:sens}).

\begin{figure}[htbp]
\begin{center}
\begin{tikzpicture}
\begin{groupplot}[group style={group size=2 by 1, horizontal sep=4.6cm},
  width=0.33\textwidth, height=6.2cm, xmode=log, log basis x=10,
  xlabel={$-E$ (MeV)}, xtick={5,10,20,50,100,200}, xticklabels={5,10,20,50,100,200},
  ymode=log, ymin=0.3, ymax=80,
  tick label style={font=\scriptsize}, label style={font=\scriptsize},
  title style={font=\scriptsize}, every mark/.append style={scale=0.8},
  legend style={font=\scriptsize, draw=none, fill=none, inner sep=1pt, row sep=-3pt}, legend cell align=left]
\nextgroupplot[title={$S$ waves}, ylabel={$\Xi(E)$ (\%)}, legend pos=outer north east]
\addplot[black, solid, mark=*] coordinates {(5,0.71) (7,0.73) (10,0.75) (14,0.76) (20,0.76) (30,0.76) (40,0.76) (50,0.76) (70,0.75) (100,0.73) (140,0.71) (200,0.68)}; \addlegendentry{$^{1}S_{0}$ $nn$}
\addplot[black, dashed, mark=square*] coordinates {(5,27.57) (7,28.72) (10,29.84) (14,30.84) (20,31.85) (30,33.03) (40,33.94) (50,34.73) (70,36.13) (100,38.01) (140,40.33) (200,43.58)}; \addlegendentry{$^{1}S_{0}$ $np$}
\addplot[black, dashdotted, mark=diamond*] coordinates {(5,1.24) (7,1.86) (10,2.58) (14,3.29) (20,4.03) (30,4.82) (40,5.31) (50,5.64) (70,6.04) (100,6.36) (140,6.55) (200,6.66)}; \addlegendentry{$^{3}S_{1}$--$^{3}D_{1}$}
\nextgroupplot[title={$np$, $j\le4$, $L\ge1$}, legend pos=outer north east]
\addplot[black, solid, mark=*] coordinates {(5,2.05) (7,2.05) (10,2.04) (14,2.04) (20,2.03) (30,2.03) (40,2.03) (50,2.03) (70,2.03) (100,2.04) (140,2.05) (200,2.06)}; \addlegendentry{$^{3}P_{0}$}
\addplot[black, dashed, mark=square*] coordinates {(5,0.63) (7,0.63) (10,0.62) (14,0.62) (20,0.61) (30,0.59) (40,0.58) (50,0.57) (70,0.55) (100,0.53) (140,0.50) (200,0.46)}; \addlegendentry{$^{3}P_{1}$}
\addplot[black, dotted, mark=triangle*] coordinates {(5,2.49) (7,2.48) (10,2.47) (14,2.45) (20,2.43) (30,2.39) (40,2.37) (50,2.34) (70,2.31) (100,2.26) (140,2.21) (200,2.16)}; \addlegendentry{$^{1}P_{1}$}
\addplot[black, dashdotted, mark=diamond*] coordinates {(5,3.46) (7,3.48) (10,3.50) (14,3.53) (20,3.56) (30,3.60) (40,3.64) (50,3.67) (70,3.72) (100,3.78) (140,3.85) (200,3.91)}; \addlegendentry{$^{3}P_{2}$--$^{3}F_{2}$}
\addplot[black, densely dashed, mark=pentagon*] coordinates {(5,1.59) (7,1.59) (10,1.59) (14,1.59) (20,1.59) (30,1.60) (40,1.60) (50,1.60) (70,1.60) (100,1.60) (140,1.60) (200,1.60)}; \addlegendentry{$^{1}D_{2}$}
\addplot[black, densely dotted, mark=o] coordinates {(5,8.50) (7,8.52) (10,8.56) (14,8.60) (20,8.65) (30,8.72) (40,8.79) (50,8.84) (70,8.92) (100,9.00) (140,9.08) (200,9.16)}; \addlegendentry{$^{3}D_{2}$}
\addplot[black, loosely dashed, mark=square] coordinates {(5,12.26) (7,12.26) (10,12.26) (14,12.26) (20,12.26) (30,12.26) (40,12.25) (50,12.24) (70,12.23) (100,12.21) (140,12.18) (200,12.15)}; \addlegendentry{$^{3}D_{3}$--$^{3}G_{3}$}
\addplot[black, loosely dotted, mark=triangle] coordinates {(5,6.17) (7,6.17) (10,6.17) (14,6.17) (20,6.17) (30,6.16) (40,6.16) (50,6.16) (70,6.15) (100,6.14) (140,6.14) (200,6.13)}; \addlegendentry{$^{3}F_{3}$}
\addplot[black, dashdotdotted, mark=diamond] coordinates {(5,6.66) (7,6.66) (10,6.66) (14,6.65) (20,6.65) (30,6.64) (40,6.63) (50,6.63) (70,6.62) (100,6.60) (140,6.58) (200,6.55)}; \addlegendentry{$^{1}F_{3}$}
\addplot[black, solid, mark=x] coordinates {(5,10.42) (7,10.43) (10,10.44) (14,10.45) (20,10.47) (30,10.49) (40,10.52) (50,10.53) (70,10.56) (100,10.59) (140,10.63) (200,10.66)}; \addlegendentry{$^{3}F_{4}$--$^{3}H_{4}$}
\end{groupplot}
\end{tikzpicture}
\end{center}
\caption{The off-shell measure $\Xi(E)$ of criterion C1 of the $NN$ separable potentials at twelve energies, evaluated from the tabulated blocks against the exact CD~Bonn $t$~matrix: the $nn$ and $np$ $^{1}S_{0}$ potentials (the $pp$ curve lies on the $np$ one within the line width) and the $^{3}S_{1}$--$^{3}D_{1}$ representation of Table~\ref{tab:eta} (left), and the $np$ potentials of the higher waves (right; the $nn$ and $pp$ members of each $T=1$ wave lie within $1.5\%$ of the $np$ one).}
\label{fig:c1nn}
\end{figure}

\begin{table}[htbp]
\caption{\label{tab:s0thr}The rank-$7$ $^{1}S_{0}$ $nn$ separable potential, built on negative-energy supports and one support at $0.15$~MeV, against the original potential. Percentages in parentheses are deviations from the original potential.}
\begin{center}
\begin{tabular}{lcc}
\toprule
quantity & separable & original \\
\midrule
$a_{nn}$ (fm) & $-18.9633$ ($0.02\%$) & $-18.9680$ \\
$r_{nn}$ (fm) & $2.8112$ ($0.25\%$) & $2.8182$ \\
$\Xi$ (\%) & $0.68$--$0.76$ & --- \\
max.\ phase deviation & $0.45^{\circ}$ & --- \\
\bottomrule
\end{tabular}
\end{center}
\end{table}  
C3 must therefore be recomputed from the fitted separable potential for every representation whose supports are optimized at $E<0$, since no other criterion protects the threshold. The rank-$7$ representation is the adopted $^{1}S_{0}$ $nn$ potential; a rank-$5$ $nn$ fit of the $np$/$pp$ construction gives practically the same $^{6}$He energy (Table~\ref{tab:sens}).

\section{The $\alpha N$ interaction}
\label{sec:an}

The KKNN interaction~\cite{KKNN1979} is a local potential built to reproduce the Coulomb-free $N\alpha$ phase shifts of the microscopic resonating-group calculation of the same authors, which in turn reproduces the empirical $n\alpha$ phases. The proton channel enters only through the Coulomb interaction added to the same nuclear potential.  Its $n\alpha$ observables therefore verify the construction, while its $p\alpha$ observables are predictions of the model. The distinction matters for the three-way comparison. The $^{5}$Li resonance below is the one $p\alpha$ prediction of the model tested here.

\subsection{Phases and the forbidden state}
\label{sec:an-phases}

The fitted separable potentials reproduce the phase shifts of the original potential over $1$--$22$~MeV very well in all five channels ($^{2}S_{1/2}$, $^{2}P_{3/2}$, $^{2}P_{1/2}$, $^{2}D_{5/2}$, $^{2}D_{3/2}$), the fine-grid comparison of criterion C2 (Fig.~\ref{fig:anphases}). The off-shell measure C1 of these representations is large: $1$--$12\%$ in $^{2}S_{1/2}$, the one wave with a bound support, and $13$--$47\%$ in the $P$ and $D$ waves, whose supports lie at positive energies only, exact on shell and unconstrained at the negative pair energies the kernel samples (Fig.~\ref{fig:c1cluster}, which shows $\Xi(E)$ for the $\alpha N$ and $\alpha\alpha$ blocks side by side); their three-body consequence is indicated by the $52$~keV spread of the $^{6}$Li energy between two accepted $\alpha N$ support sets (Table~\ref{tab:sens}). The $^{2}S_{1/2}$ Pauli-forbidden state of the fitted representation agrees closely with the value of the original potential ($-12.264$~MeV), satisfying C4.

\begin{figure}[htbp]
\begin{center}
\begin{tikzpicture}
\begin{axis}[width=0.8\textwidth, height=6.4cm, xlabel={$E_{\rm cm}$ (MeV)}, ylabel={$\delta$ (deg)},
  xmin=0, xmax=23, tick label style={font=\scriptsize}, label style={font=\scriptsize},
  legend style={font=\scriptsize, draw=none, fill=none, inner sep=1pt, row sep=-3pt}, legend cell align=left,
  legend pos=outer north east, every mark/.append style={scale=0.9}]
\addplot[black, solid, forget plot] coordinates {(0.25,165.212) (0.5,159.171) (0.75,154.593) (1,150.780) (1.25,147.460) (1.5,144.493) (1.75,141.797) (2,139.317) (2.25,137.014) (2.5,134.860) (2.75,132.835) (3,130.922) (3.25,129.107) (3.5,127.380) (3.75,125.732) (4,124.156) (4.25,122.645) (4.5,121.194) (4.75,119.797) (5,118.452) (5.25,117.154) (5.5,115.901) (5.75,114.688) (6,113.514) (6.25,112.377) (6.5,111.274) (6.75,110.204) (7,109.164) (7.25,108.154) (7.5,107.171) (7.75,106.215) (8,105.284) (8.25,104.376) (8.5,103.492) (8.75,102.630) (9,101.789) (9.25,100.968) (9.5,100.167) (9.75,99.385) (10,98.620) (10.25,97.873) (10.5,97.142) (10.75,96.428) (11,95.729) (11.25,95.045) (11.5,94.375) (11.75,93.720) (12,93.078) (12.25,92.450) (12.5,91.834) (12.75,91.230) (13,90.639) (13.25,90.059) (13.5,89.491) (13.75,88.933) (14,88.386) (14.25,87.850) (14.5,87.323) (14.75,86.807) (15,86.299) (15.25,85.802) (15.5,85.313) (15.75,84.833) (16,84.361) (16.25,83.898) (16.5,83.443) (16.75,82.996) (17,82.557) (17.25,82.125) (17.5,81.700) (17.75,81.283) (18,80.873) (18.25,80.469) (18.5,80.072) (18.75,79.682) (19,79.298) (19.25,78.920) (19.5,78.548) (19.75,78.182) (20,77.822) (20.25,77.468) (20.5,77.119) (20.75,76.776) (21,76.437) (21.25,76.104) (21.5,75.776) (21.75,75.453) (22,75.135)};
\addplot[black, only marks, mark=*] coordinates {(0.5,159.17) (1,150.78) (2,139.32) (3,130.92) (5,118.45) (8,105.28) (12,93.06) (16,84.34) (22,75.13)}; \addlegendentry{$^{2}S_{1/2}$}
\addplot[black, dashed, forget plot] coordinates {(0.25,5.980) (0.5,28.373) (0.75,78.632) (1,110.744) (1.25,122.003) (1.5,126.378) (1.75,128.152) (2,128.738) (2.25,128.711) (2.5,128.338) (2.75,127.761) (3,127.059) (3.25,126.281) (3.5,125.456) (3.75,124.604) (4,123.738) (4.25,122.866) (4.5,121.995) (4.75,121.129) (5,120.271) (5.25,119.423) (5.5,118.586) (5.75,117.761) (6,116.950) (6.25,116.151) (6.5,115.366) (6.75,114.594) (7,113.835) (7.25,113.090) (7.5,112.358) (7.75,111.639) (8,110.933) (8.25,110.239) (8.5,109.557) (8.75,108.887) (9,108.229) (9.25,107.582) (9.5,106.946) (9.75,106.321) (10,105.706) (10.25,105.101) (10.5,104.507) (10.75,103.922) (11,103.347) (11.25,102.781) (11.5,102.224) (11.75,101.675) (12,101.135) (12.25,100.604) (12.5,100.081) (12.75,99.565) (13,99.058) (13.25,98.557) (13.5,98.065) (13.75,97.579) (14,97.100) (14.25,96.629) (14.5,96.164) (14.75,95.705) (15,95.253) (15.25,94.807) (15.5,94.367) (15.75,93.933) (16,93.505) (16.25,93.083) (16.5,92.666) (16.75,92.255) (17,91.849) (17.25,91.448) (17.5,91.052) (17.75,90.661) (18,90.275) (18.25,89.894) (18.5,89.518) (18.75,89.146) (19,88.779) (19.25,88.416) (19.5,88.057) (19.75,87.703) (20,87.352) (20.25,87.006) (20.5,86.664) (20.75,86.326) (21,85.991) (21.25,85.661) (21.5,85.334) (21.75,85.010) (22,84.690)};
\addplot[black, only marks, mark=square*] coordinates {(0.5,28.43) (1,110.83) (2,128.76) (3,127.07) (5,120.28) (8,110.95) (12,101.16) (16,93.52) (22,84.71)}; \addlegendentry{$^{2}P_{3/2}$}
\addplot[black, dotted, forget plot] coordinates {(0.25,0.709) (0.5,2.053) (0.75,3.851) (1,6.036) (1.25,8.557) (1.5,11.362) (1.75,14.394) (2,17.589) (2.25,20.880) (2.5,24.196) (2.75,27.472) (3,30.649) (3.25,33.681) (3.5,36.531) (3.75,39.178) (4,41.608) (4.25,43.819) (4.5,45.815) (4.75,47.605) (5,49.200) (5.25,50.616) (5.5,51.866) (5.75,52.965) (6,53.926) (6.25,54.764) (6.5,55.490) (6.75,56.115) (7,56.650) (7.25,57.104) (7.5,57.485) (7.75,57.801) (8,58.058) (8.25,58.263) (8.5,58.420) (8.75,58.534) (9,58.611) (9.25,58.652) (9.5,58.663) (9.75,58.645) (10,58.603) (10.25,58.537) (10.5,58.451) (10.75,58.346) (11,58.224) (11.25,58.087) (11.5,57.936) (11.75,57.773) (12,57.599) (12.25,57.414) (12.5,57.221) (12.75,57.019) (13,56.810) (13.25,56.594) (13.5,56.372) (13.75,56.145) (14,55.913) (14.25,55.677) (14.5,55.437) (14.75,55.194) (15,54.948) (15.25,54.699) (15.5,54.448) (15.75,54.195) (16,53.940) (16.25,53.684) (16.5,53.427) (16.75,53.169) (17,52.910) (17.25,52.651) (17.5,52.391) (17.75,52.131) (18,51.871) (18.25,51.611) (18.5,51.351) (18.75,51.091) (19,50.832) (19.25,50.573) (19.5,50.315) (19.75,50.057) (20,49.800) (20.25,49.544) (20.5,49.289) (20.75,49.035) (21,48.782) (21.25,48.529) (21.5,48.278) (21.75,48.028) (22,47.779)};
\addplot[black, only marks, mark=triangle*] coordinates {(0.5,2.05) (1,6.03) (2,17.55) (3,30.61) (5,49.21) (8,58.04) (12,57.53) (16,53.90) (22,47.79)}; \addlegendentry{$^{2}P_{1/2}$}
\addplot[black, dashdotted, forget plot] coordinates {(0.25,179.999) (0.5,179.997) (0.75,179.991) (1,179.982) (1.25,179.970) (1.5,179.954) (1.75,179.935) (2,179.913) (2.25,179.888) (2.5,179.859) (2.75,179.829) (3,179.796) (3.25,179.761) (3.5,179.724) (3.75,179.686) (4,179.647) (4.25,179.607) (4.5,179.566) (4.75,179.525) (5,179.483) (5.25,179.442) (5.5,179.402) (5.75,179.362) (6,179.324) (6.25,179.286) (6.5,179.251) (6.75,179.217) (7,179.185) (7.25,179.155) (7.5,179.128) (7.75,179.104) (8,179.082) (8.25,179.064) (8.5,179.049) (8.75,179.037) (9,179.029) (9.25,179.025) (9.5,179.025) (9.75,179.029) (10,179.037) (10.25,179.050) (10.5,179.068) (10.75,179.090) (11,179.117) (11.25,179.149) (11.5,179.186) (11.75,179.229) (12,179.277) (12.25,179.330) (12.5,179.389) (12.75,179.454) (13,179.524) (13.25,179.600) (13.5,179.682) (13.75,179.771) (14,179.865) (14.25,179.965) (14.5,180.072) (14.75,180.185) (15,180.304) (15.25,180.429) (15.5,180.561) (15.75,180.700) (16,180.845) (16.25,180.996) (16.5,181.154) (16.75,181.319) (17,181.490) (17.25,181.668) (17.5,181.852) (17.75,182.043) (18,182.241) (18.25,182.445) (18.5,182.656) (18.75,182.873) (19,183.097) (19.25,183.327) (19.5,183.564) (19.75,183.807) (20,184.056) (20.25,184.312) (20.5,184.574) (20.75,184.842) (21,185.116) (21.25,185.397) (21.5,185.683) (21.75,185.975) (22,186.273)};
\addplot[black, only marks, mark=diamond*] coordinates {(0.5,180.00) (1,179.98) (2,179.91) (3,179.80) (5,179.48) (8,179.09) (12,179.28) (16,180.85) (22,186.26)}; \addlegendentry{$^{2}D_{5/2}$}
\addplot[black, densely dotted, forget plot] coordinates {(0.25,179.999) (0.5,179.995) (0.75,179.987) (1,179.975) (1.25,179.957) (1.5,179.935) (1.75,179.907) (2,179.874) (2.25,179.836) (2.5,179.793) (2.75,179.745) (3,179.693) (3.25,179.636) (3.5,179.576) (3.75,179.511) (4,179.443) (4.25,179.371) (4.5,179.297) (4.75,179.220) (5,179.140) (5.25,179.057) (5.5,178.973) (5.75,178.887) (6,178.799) (6.25,178.710) (6.5,178.619) (6.75,178.528) (7,178.436) (7.25,178.343) (7.5,178.250) (7.75,178.157) (8,178.064) (8.25,177.971) (8.5,177.878) (8.75,177.786) (9,177.694) (9.25,177.603) (9.5,177.513) (9.75,177.424) (10,177.337) (10.25,177.250) (10.5,177.165) (10.75,177.081) (11,176.999) (11.25,176.919) (11.5,176.840) (11.75,176.763) (12,176.688) (12.25,176.615) (12.5,176.544) (12.75,176.475) (13,176.409) (13.25,176.344) (13.5,176.282) (13.75,176.222) (14,176.165) (14.25,176.110) (14.5,176.057) (14.75,176.007) (15,175.959) (15.25,175.914) (15.5,175.872) (15.75,175.832) (16,175.795) (16.25,175.760) (16.5,175.728) (16.75,175.699) (17,175.672) (17.25,175.648) (17.5,175.627) (17.75,175.609) (18,175.593) (18.25,175.580) (18.5,175.569) (18.75,175.561) (19,175.556) (19.25,175.554) (19.5,175.554) (19.75,175.557) (20,175.563) (20.25,175.571) (20.5,175.582) (20.75,175.596) (21,175.612) (21.25,175.631) (21.5,175.652) (21.75,175.676) (22,175.702) (22.25,175.731) (22.5,175.763) (22.75,175.797) (23,175.833)};
\addplot[black, only marks, mark=o] coordinates {(0.5,180.00) (1,179.97) (2,179.87) (3,179.69) (5,179.14) (8,178.06) (12,176.69) (16,175.79) (22,175.70)}; \addlegendentry{$^{2}D_{3/2}$}
\end{axis}
\end{tikzpicture}
\end{center}
\caption{The $\alpha n$ phase shifts of the KKNN potential (lines) and of the fitted separable potentials (points), criterion C2; the $^{2}S_{1/2}$ phase is the unprojected one, starting at $180^{\circ}$ by Levinson's theorem for the forbidden state, and the phases are drawn continuous.}
\label{fig:anphases}
\end{figure}
\begin{figure}[htbp]
\begin{center}
\begin{tikzpicture}
\begin{groupplot}[group style={group size=2 by 1, horizontal sep=1.6cm},
  width=0.5\textwidth, height=5.6cm, xmode=log, log basis x=10,
  xlabel={$-E$ (MeV)}, xtick={5,10,20,50,100,200}, xticklabels={5,10,20,50,100,200},
  ymode=log, ymin=0.15, ymax=60,
  tick label style={font=\scriptsize}, label style={font=\scriptsize},
  title style={font=\scriptsize}, every mark/.append style={scale=0.9},
  legend style={font=\scriptsize, draw=none, fill=none, inner sep=1pt, row sep=-3pt}, legend cell align=left]
\nextgroupplot[title={$\alpha n$}, ylabel={$\Xi(E)$ (\%)}, legend pos=south east]
\addplot[black, solid, mark=*] coordinates {(5,6.90) (7,4.40) (10,1.60) (14,1.04) (20,3.52) (30,5.93) (40,7.34) (50,8.27) (70,9.43) (100,10.38) (140,11.06) (200,11.59)}; \addlegendentry{$^{2}S_{1/2}$}
\addplot[black, dashed, mark=square*] coordinates {(5,23.78) (7,26.76) (10,29.99) (14,33.00) (20,36.03) (30,39.15) (40,41.08) (50,42.41) (70,44.11) (100,45.53) (140,46.53) (200,47.31)}; \addlegendentry{$^{2}P_{3/2}$}
\addplot[black, dotted, mark=triangle*] coordinates {(5,12.63) (7,13.16) (10,13.78) (14,14.40) (20,15.10) (30,15.91) (40,16.50) (50,16.97) (70,17.72) (100,18.59) (140,19.56) (200,20.79)}; \addlegendentry{$^{2}P_{1/2}$}
\addplot[black, dashdotted, mark=diamond*] coordinates {(5,28.63) (7,28.84) (10,29.11) (14,29.43) (20,29.82) (30,30.32) (40,30.69) (50,30.97) (70,31.38) (100,31.74) (140,32.01) (200,32.20)}; \addlegendentry{$^{2}D_{5/2}$}
\addplot[black, densely dashed, mark=pentagon*] coordinates {(5,20.77) (7,20.79) (10,20.82) (14,20.85) (20,20.88) (30,20.91) (40,20.92) (50,20.92) (70,20.91) (100,20.87) (140,20.80) (200,20.67)}; \addlegendentry{$^{2}D_{3/2}$}
\nextgroupplot[title={$\alpha\alpha$}, legend pos=north west]
\addplot[black, solid, mark=*] coordinates {(5,1.46) (7,1.06) (10,0.94) (14,0.80) (20,0.68) (30,4.27) (40,0.83) (50,1.32) (70,3.03) (100,1.76) (140,1.18) (200,1.39)}; \addlegendentry{$L=0$}
\addplot[black, dashed, mark=square*] coordinates {(5,1.40) (7,1.49) (10,1.35) (14,0.97) (20,0.42) (30,0.24) (40,0.71) (50,1.07) (70,1.59) (100,2.13) (140,2.60) (200,3.04)}; \addlegendentry{$L=2$}
\addplot[black, dotted, mark=triangle*] coordinates {(5,1.16) (7,1.24) (10,1.35) (14,1.49) (20,1.66) (30,1.89) (40,2.08) (50,2.24) (70,2.49) (100,2.75) (140,3.00) (200,3.24)}; \addlegendentry{$L=4$}
\end{groupplot}
\end{tikzpicture}
\end{center}
\caption{The off-shell measure $\Xi(E)$ of criterion C1, Eq.~(\ref{eq:eps}), of the cluster separable potentials at twelve energies from $-5$ to $-200$~MeV, evaluated from the tabulated blocks against the exact $t$~matrix of the original potential with the Coulomb interaction off: the five $\alpha n$ waves (left) and the three $\alpha\alpha$ waves (right).  The two bumps of the $\alpha\alpha$ $L=0$ curve at $-30$ and $-70$~MeV lie on the flanks of the poles of its two Pauli-forbidden bound states, which with the Coulomb interaction subtracted are at $-31.1$ and $-80.3$~MeV, where the relative deviation of two nearly singular amplitudes is magnified.}
\label{fig:c1cluster}
\end{figure}

\subsection{Resonances from the fitted separable potentials}
\label{sec:an-res}

Resonance parameters are extracted from the fitted separable potentials by analytic continuation: the quantity

\begin{equation}
\det\bm{\tau}^{-1}(E)=\det[\bm{\Lambda}^{-1}-\bm{\mathcal D}(E)]
\label{eq:taudet}
\end{equation}
is computed on a sub-threshold grid and fitted by a rational (Pad\'e) approximant in the momentum variable $k$, $E=k^{2}/2\mu$, in which the two Riemann sheets of the energy plane unfold, and the resonance is the zero $k_{0}$ of $\det\bm{\tau}^{-1}$ in the fourth quadrant of the $k$~plane,

\begin{equation}
E_{r}-i\Gamma/2=k_{0}^{2}/2\mu.
\label{eq:pole}
\end{equation}
The precision of the method, estimated from the stability across Pad\'e orders and against an independent coordinate-space determination, is a small fraction of a percent on the broad states considered here. The $^{5}$Li values are extrapolated in $R$ from $R=50$--$200$~fm (Sec.~\ref{sec:coulomb}); the position moves by a few tens of keV over this range. Figure~\ref{fig:apphases} shows the Coulomb-modified $\alpha p$ phases themselves, the comparison of Fig.~\ref{fig:anphases} with the Coulomb interaction present and treated exactly on both sides: the fitted separable potentials follow the original potential through the $^{5}$Li resonance to the method precision.

\begin{figure}[htbp]
\begin{center}
\begin{tikzpicture}
\begin{groupplot}[group style={group size=3 by 2, horizontal sep=1.15cm, vertical sep=1.6cm},
  width=0.33\textwidth, height=4.4cm, xmin=0, xmax=23, xlabel={$E_{\rm cm}$ (MeV)}, ylabel={$\delta_{SC}$ (deg)},
  tick label style={font=\scriptsize}, label style={font=\scriptsize}, title style={font=\scriptsize, yshift=-1ex}]
\nextgroupplot[title={$^{2}S_{1/2}$}]
\addplot[black, solid, forget plot] coordinates {(0.5,-8.046) (1,-16.833) (1.5,-23.854) (2,-29.720) (2.5,-34.785) (3,-39.258) (3.5,-43.273) (4,-46.920) (4.5,-50.263) (5,-53.350) (5.5,-56.217) (6,-58.895) (6.5,-61.404) (7,-63.764) (7.5,-65.992) (8,-68.099) (8.5,-70.096) (9,-71.994) (9.5,-73.801) (10,-75.524) (10.5,-77.168) (11,-78.740) (11.5,-80.245) (12,-81.688) (12.5,-83.071) (13,-84.399) (13.5,-85.676) (14,-86.904) (14.5,-88.085) (15,-89.223) (15.5,-90.320) (16,-91.378) (16.5,-92.399) (17,-93.385) (17.5,-94.337) (18,-95.258) (18.5,-96.148) (19,-97.009) (19.5,-97.843) (20,-98.651) (20.5,-99.433) (21,-100.191) (21.5,-100.927) (22,-101.640)};
\addplot[only marks, mark=*, mark size=1.1pt, forget plot] coordinates {(0.5,-8.047) (1.5,-23.855) (2.5,-34.786) (3.5,-43.276) (4.5,-50.266) (5.5,-56.222) (6.5,-61.409) (7.5,-65.998) (8.5,-70.103) (9.5,-73.809) (10.5,-77.178) (11.5,-80.257) (12.5,-83.084) (13.5,-85.690) (14.5,-88.099) (15.5,-90.334) (16.5,-92.411) (17.5,-94.346) (18.5,-96.153) (19.5,-97.843) (20.5,-99.427) (21.5,-100.914)};
\nextgroupplot[title={$^{2}P_{3/2}$}]
\addplot[black, solid, forget plot] coordinates {(0.25,0.389) (0.5,2.369) (0.75,6.765) (1,14.968) (1.25,29.132) (1.5,50.307) (1.75,73.409) (2,90.966) (2.25,101.959) (2.5,108.570) (2.75,112.592) (3,115.068) (3.25,116.583) (3.5,117.473) (3.75,117.942) (4,118.120) (4.25,118.090) (4.5,117.909) (4.75,117.617) (5,117.243) (5.25,116.806) (5.5,116.323) (5.75,115.805) (6,115.260) (6.25,114.697) (6.5,114.119) (6.75,113.531) (7,112.937) (7.25,112.339) (7.5,111.739) (7.75,111.139) (8,110.540) (8.25,109.943) (8.5,109.350) (8.75,108.761) (9,108.176) (9.25,107.597) (9.5,107.022) (9.75,106.454) (10,105.891) (10.25,105.334) (10.5,104.784) (10.75,104.240) (11,103.702) (11.25,103.171) (11.5,102.646) (11.75,102.127) (12,101.615) (12.25,101.109) (12.5,100.609) (12.75,100.115) (13,99.628) (13.25,99.146) (13.5,98.671) (13.75,98.201) (14,97.738) (14.25,97.280) (14.5,96.828) (14.75,96.381) (15,95.940) (15.25,95.504) (15.5,95.074) (15.75,94.648) (16,94.228) (16.25,93.813) (16.5,93.403) (16.75,92.998) (17,92.598) (17.25,92.202) (17.5,91.811) (17.75,91.425) (18,91.043) (18.25,90.665) (18.5,90.292) (18.75,89.923) (19,89.558) (19.25,89.197) (19.5,88.841) (19.75,88.488) (20,88.139) (20.25,87.794) (20.5,87.453) (20.75,87.115) (21,86.782) (21.25,86.451) (21.5,86.124) (21.75,85.801) (22,85.481)};
\addplot[only marks, mark=*, mark size=1.1pt, forget plot] coordinates {(0.25,0.391) (1.25,29.166) (2.25,101.928) (3.25,116.560) (4.25,118.056) (5.25,116.782) (6.25,114.667) (7.25,112.312) (8.25,109.909) (9.25,107.565) (10.25,105.292) (11.25,103.127) (12.25,101.058) (13.25,99.086) (14.25,97.218) (15.25,95.437) (16.25,93.740) (17.25,92.131) (18.25,90.596) (19.25,89.127) (20.25,87.730) (21.25,86.396)};
\nextgroupplot[title={$^{2}P_{1/2}$}]
\addplot[black, solid, forget plot] coordinates {(0.25,0.104) (0.5,0.553) (0.75,1.341) (1,2.440) (1.25,3.828) (1.5,5.486) (1.75,7.392) (2,9.524) (2.25,11.851) (2.5,14.342) (2.75,16.956) (3,19.655) (3.25,22.393) (3.5,25.131) (3.75,27.827) (4,30.447) (4.25,32.963) (4.5,35.351) (4.75,37.596) (5,39.688) (5.25,41.623) (5.5,43.399) (5.75,45.021) (6,46.494) (6.25,47.826) (6.5,49.024) (6.75,50.099) (7,51.059) (7.25,51.912) (7.5,52.668) (7.75,53.335) (8,53.920) (8.25,54.431) (8.5,54.874) (8.75,55.255) (9,55.580) (9.25,55.853) (9.5,56.080) (9.75,56.265) (10,56.410) (10.25,56.521) (10.5,56.599) (10.75,56.649) (11,56.671) (11.25,56.670) (11.5,56.646) (11.75,56.602) (12,56.539) (12.25,56.460) (12.5,56.365) (12.75,56.256) (13,56.134) (13.25,56.001) (13.5,55.857) (13.75,55.703) (14,55.540) (14.25,55.370) (14.5,55.191) (14.75,55.007) (15,54.816) (15.25,54.619) (15.5,54.418) (15.75,54.212) (16,54.001) (16.25,53.788) (16.5,53.570) (16.75,53.350) (17,53.127) (17.25,52.902) (17.5,52.675) (17.75,52.446) (18,52.215) (18.25,51.983) (18.5,51.749) (18.75,51.515) (19,51.280) (19.25,51.044) (19.5,50.807) (19.75,50.570) (20,50.333) (20.25,50.096) (20.5,49.859) (20.75,49.622) (21,49.384) (21.25,49.148) (21.5,48.911) (21.75,48.675) (22,48.439)};
\addplot[only marks, mark=*, mark size=1.1pt, forget plot] coordinates {(0.25,0.104) (1.25,3.819) (2.25,11.835) (3.25,22.393) (4.25,32.988) (5.25,41.654) (6.25,47.831) (7.25,51.876) (8.25,54.350) (9.25,55.737) (10.25,56.379) (11.25,56.517) (12.25,56.307) (13.25,55.857) (14.25,55.243) (15.25,54.512) (16.25,53.701) (17.25,52.838) (18.25,51.937) (19.25,51.011) (20.25,50.074) (21.25,49.132)};
\nextgroupplot[title={$^{2}D_{5/2}$}]
\addplot[black, solid, forget plot] coordinates {(0.25,0.000) (0.5,-0.001) (0.75,-0.004) (1,-0.009) (1.25,-0.016) (1.5,-0.027) (1.75,-0.040) (2,-0.056) (2.25,-0.075) (2.5,-0.097) (2.75,-0.122) (3,-0.149) (3.25,-0.179) (3.5,-0.211) (3.75,-0.244) (4,-0.280) (4.25,-0.317) (4.5,-0.355) (4.75,-0.394) (5,-0.434) (5.25,-0.474) (5.5,-0.514) (5.75,-0.555) (6,-0.595) (6.25,-0.635) (6.5,-0.673) (6.75,-0.711) (7,-0.748) (7.25,-0.783) (7.5,-0.817) (7.75,-0.848) (8,-0.878) (8.25,-0.905) (8.5,-0.930) (8.75,-0.952) (9,-0.971) (9.25,-0.988) (9.5,-1.001) (9.75,-1.010) (10,-1.016) (10.25,-1.019) (10.5,-1.017) (10.75,-1.012) (11,-1.003) (11.25,-0.989) (11.5,-0.971) (11.75,-0.948) (12,-0.921) (12.25,-0.889) (12.5,-0.852) (12.75,-0.810) (13,-0.763) (13.25,-0.711) (13.5,-0.653) (13.75,-0.591) (14,-0.523) (14.25,-0.449) (14.5,-0.370) (14.75,-0.285) (15,-0.194) (15.25,-0.098) (15.5,0.005) (15.75,0.113) (16,0.227) (16.25,0.347) (16.5,0.473) (16.75,0.606) (17,0.744) (17.25,0.889) (17.5,1.039) (17.75,1.196) (18,1.359) (18.25,1.529) (18.5,1.704) (18.75,1.886) (19,2.074) (19.25,2.268) (19.5,2.468) (19.75,2.675) (20,2.887) (20.25,3.106) (20.5,3.331) (20.75,3.562) (21,3.799) (21.25,4.042) (21.5,4.291) (21.75,4.545) (22,4.806)};
\addplot[only marks, mark=*, mark size=1.1pt, forget plot] coordinates {(0.25,-0.000) (1.25,-0.016) (2.25,-0.076) (3.25,-0.179) (4.25,-0.317) (5.25,-0.474) (6.25,-0.634) (7.25,-0.783) (8.25,-0.905) (9.25,-0.987) (10.25,-1.019) (11.25,-0.989) (12.25,-0.889) (13.25,-0.712) (14.25,-0.450) (15.25,-0.099) (16.25,0.345) (17.25,0.887) (18.25,1.527) (19.25,2.268) (20.25,3.107) (21.25,4.044)};
\nextgroupplot[title={$^{2}D_{3/2}$}]
\addplot[black, solid, forget plot] coordinates {(0.25,-0.000) (0.5,-0.002) (0.75,-0.005) (1,-0.012) (1.25,-0.022) (1.5,-0.037) (1.75,-0.056) (2,-0.079) (2.25,-0.106) (2.5,-0.139) (2.75,-0.175) (3,-0.216) (3.25,-0.262) (3.5,-0.311) (3.75,-0.364) (4,-0.422) (4.25,-0.482) (4.5,-0.546) (4.75,-0.614) (5,-0.684) (5.25,-0.757) (5.5,-0.832) (5.75,-0.910) (6,-0.990) (6.25,-1.071) (6.5,-1.155) (6.75,-1.240) (7,-1.326) (7.25,-1.413) (7.5,-1.501) (7.75,-1.589) (8,-1.679) (8.25,-1.768) (8.5,-1.858) (8.75,-1.948) (9,-2.037) (9.25,-2.127) (9.5,-2.215) (9.75,-2.304) (10,-2.391) (10.25,-2.478) (10.5,-2.564) (10.75,-2.649) (11,-2.733) (11.25,-2.815) (11.5,-2.896) (11.75,-2.976) (12,-3.054) (12.25,-3.131) (12.5,-3.206) (12.75,-3.279) (13,-3.350) (13.25,-3.420) (13.5,-3.487) (13.75,-3.553) (14,-3.616) (14.25,-3.677) (14.5,-3.736) (14.75,-3.793) (15,-3.848) (15.25,-3.900) (15.5,-3.950) (15.75,-3.998) (16,-4.043) (16.25,-4.086) (16.5,-4.126) (16.75,-4.164) (17,-4.199) (17.25,-4.232) (17.5,-4.262) (17.75,-4.290) (18,-4.316) (18.25,-4.338) (18.5,-4.359) (18.75,-4.376) (19,-4.391) (19.25,-4.404) (19.5,-4.414) (19.75,-4.421) (20,-4.426) (20.25,-4.428) (20.5,-4.428) (20.75,-4.425) (21,-4.419) (21.25,-4.411) (21.5,-4.401) (21.75,-4.388) (22,-4.373)};
\addplot[only marks, mark=*, mark size=1.1pt, forget plot] coordinates {(0.25,-0.000) (1.25,-0.023) (2.25,-0.107) (3.25,-0.262) (4.25,-0.482) (5.25,-0.755) (6.25,-1.069) (7.25,-1.410) (8.25,-1.766) (9.25,-2.125) (10.25,-2.478) (11.25,-2.816) (12.25,-3.132) (13.25,-3.421) (14.25,-3.678) (15.25,-3.901) (16.25,-4.086) (17.25,-4.233) (18.25,-4.339) (19.25,-4.406) (20.25,-4.432) (21.25,-4.417)};
\end{groupplot}
\end{tikzpicture}
\end{center}
\caption{The Coulomb-modified nuclear phase shifts $\delta_{SC}$ of the $\alpha p$ system, criterion C2 with the Coulomb interaction present: the KKNN potential with the point Coulomb force (lines) and the fitted separable potentials with the same Coulomb force treated exactly (points, every second grid point in $^{2}S_{1/2}$ and every fourth in the $P$ and $D$ waves), in all five waves. The phases are drawn continuous on the branch with $\delta_{SC}=0$ at threshold, the $^{2}S_{1/2}$ phase being the one with the forbidden state projected. Fit and original potential agree to a fraction of a degree in every wave, including the region of the $^{5}$Li $3/2^{-}$ resonance, where the $^{2}P_{3/2}$ phase rises steeply.}
\label{fig:apphases}
\end{figure}

\begin{table}[htbp]
\caption{\label{tab:anres}$A=5$ resonances (MeV): fit and original potential agree to the method precision, so a single column is shown for both, against the experimental compilation of Ref.~\cite{Tilley2002}; the last column is the relative deviation of the fit from experiment, $(\mathrm{fit}-\mathrm{exp})/\mathrm{exp}$, for $E_{r}$ and $\Gamma$ in turn; a negative entry means the fitted resonance lies lower, or is narrower, than experiment. The Coulomb-shift row is formed from the extrapolated $^{5}$Li position and the unrounded $^{5}$He position.}
\begin{center}
\begin{tabular}{lccc}
\toprule
state & fit/original & experiment & deviation (\%) \\
\midrule
$^{5}$He $3/2^{-}$: $E_{r}$, $\Gamma$ & $0.668$, $0.496$ & $0.798$, $0.648$ & $-16$, $-23$ \\
$^{5}$He $1/2^{-}$: $E_{r}$, $\Gamma$ & $2.151$, $6.10$ & $2.07$, $5.57$ & $+4$, $+10$ \\
$^{5}$Li $3/2^{-}$: $E_{r}$, $\Gamma$ & $1.55$, $1.12$   & $1.69$, $1.23$   & $-8$, $-9$ \\
\midrule
$^{5}\mathrm{He}\!\to\!{}^{5}\mathrm{Li}$ shift & $0.88$ & $0.89$ & $-1$ \\
\bottomrule
\end{tabular}
\end{center}
\end{table}

The results are collected in Table~\ref{tab:anres}. The narrow $3/2^{-}$ ground-state resonance of $^{5}$He comes out low in position and narrow relative to experiment. This is a property of KKNN itself, since the fitted separable potential, the Pad\'e continuation, and a coordinate-space solution of the same Hamiltonian reproduce it to the method precision. The broad $1/2^{-}$ and the $p\alpha$ prediction, the $^{5}$Li $3/2^{-}$ resonance, agree with experiment to within about ten percent. The most informative entry is the $^{5}\mathrm{He}\to{}^{5}\mathrm{Li}$ Coulomb shift, in which the common deficit of the model cancels.  It is reproduced to a few percent, which shows the Coulomb strength and the kinematics of the $\alpha p$ system to be consistent, independently of the quality of the nuclear model.

The deficit of the $3/2^{-}$ position has a quantifiable three-body consequence. Scaling the $^{2}P_{3/2}$ strength by $0.983$ moves the resonance onto the experimental position. The corresponding separable potential changes $E(^{6}\mathrm{Li})$ by $+0.29$~MeV (less bound). The low resonance position of the model thus masks part of the attractive three-body residual in $A=6$ but does not create it. This number bounds the part of the $A=6$ residual attributable to the resonance deficit.

\section{The $\alpha\alpha$ interaction}
\label{sec:aa}

\subsection{The potential is defined with its folded Coulomb}
\label{sec:folded}

The BFW potential~\cite{BFW1977} is a single Gaussian well fitted to $\alpha\alpha$ phases \emph{together with} the folded Coulomb interaction~(\ref{eq:folded}) at $Z_{1}Z_{2}=4$, $\beta_{f}=0.75\,\mathrm{fm}^{-1}$, which places the $^{8}$Be $0^{+}$ resonance at $87$~keV ($92$~keV in experiment). Using a point Coulomb with BFW's nuclear parameters is therefore inconsistent, the nuclear part having absorbed the folding during the fit: the $^{8}$Be $0^{+}$ pole moves to $445$~keV and acquires a width of $78$~keV, and the Pauli-forbidden states shift by MeV (Table~\ref{tab:pointcoul}). A $^{9}$Be three-body calculation samples precisely this near-threshold amplitude (see criterion C5), so the choice of the Coulomb form enters its binding energy directly. The folding is treated exactly, the separable potentials being built from the short-range original potential

\begin{equation}
V_{\mathrm{BFW}}(r)-4e^{2}\,\frac{\mathrm{erfc}(\beta_{f}r)}{r},
\label{eq:shortrange}
\end{equation}
with the subtracted term represented as a sixteen-term Gaussian quadrature; the separable potential plus the point-Coulomb representation then reproduces the published folded interaction. The fitted separable potentials reproduce the nuclear phases $\delta_{N}$ of the BFW potential, the screened Coulomb phase subtracted, over $0.5$--$22$~MeV (Fig.~\ref{fig:aaphases}), and the Coulomb-modified phases $\delta_{SC}$ with the Coulomb interaction treated exactly on both sides to about a degree, the largest deviations lying at the top of the range rather than at the resonances (Fig.~\ref{fig:aascphases}), criterion C2.

\begin{figure}[htbp]
\begin{center}
\begin{tikzpicture}
\begin{axis}[width=0.8\textwidth, height=6.4cm, xlabel={$E_{\rm cm}$ (MeV)}, ylabel={$\delta$ (deg)},
  xmin=0, xmax=23, tick label style={font=\scriptsize}, label style={font=\scriptsize},
  legend style={font=\scriptsize, draw=none, fill=none, inner sep=1pt, row sep=-3pt}, legend cell align=left,
  legend pos=outer north east, every mark/.append style={scale=0.9}]
\addplot[black, solid, forget plot] coordinates {(0.25,178.786) (0.5,170.429) (0.75,158.960) (1,147.725) (1.25,137.454) (1.5,128.179) (1.75,119.785) (2,112.143) (2.25,105.139) (2.5,98.679) (2.75,92.687) (3,87.098) (3.25,81.864) (3.5,76.941) (3.75,72.295) (4,67.895) (4.25,63.718) (4.5,59.742) (4.75,55.948) (5,52.321) (5.25,48.847) (5.5,45.513) (5.75,42.308) (6,39.224) (6.25,36.252) (6.5,33.383) (6.75,30.612) (7,27.932) (7.25,25.337) (7.5,22.822) (7.75,20.382) (8,18.014) (8.25,15.713) (8.5,13.476) (8.75,11.300) (9,9.180) (9.25,7.115) (9.5,5.102) (9.75,3.138) (10,1.222) (10.25,-0.650) (10.5,-2.479) (10.75,-4.267) (11,-6.015) (11.25,-7.726) (11.5,-9.400) (11.75,-11.040) (12,-12.647) (12.25,-14.222) (12.5,-15.766) (12.75,-17.280) (13,-18.766) (13.25,-20.224) (13.5,-21.656) (13.75,-23.062) (14,-24.444) (14.25,-25.802) (14.5,-27.136) (14.75,-28.449) (15,-29.740) (15.25,-31.010) (15.5,-32.259) (15.75,-33.490) (16,-34.701) (16.25,-35.894) (16.5,-37.069) (16.75,-38.227) (17,-39.368) (17.25,-40.492) (17.5,-41.601) (17.75,-42.695) (18,-43.773) (18.25,-44.837) (18.5,-45.886) (18.75,-46.922) (19,-47.944) (19.25,-48.953) (19.5,-49.949) (19.75,-50.933) (20,-51.905) (20.25,-52.865) (20.5,-53.813) (20.75,-54.750) (21,-55.676) (21.25,-56.591) (21.5,-57.495) (21.75,-58.389) (22,-59.274)};
\addplot[black, only marks, mark=*] coordinates {(0.5,170.10) (1,147.10) (2,111.53) (3,86.76) (5,52.75) (8,17.58) (12,-13.49) (16,-35.46) (22,-59.98)}; \addlegendentry{$L=0$}
\addplot[black, dashed, forget plot] coordinates {(0.25,0.000) (0.5,0.019) (0.75,0.135) (1,0.486) (1.25,1.268) (1.5,2.758) (1.75,5.380) (2,9.820) (2.25,17.206) (2.5,29.186) (2.75,46.951) (3,67.951) (3.25,86.041) (3.5,98.393) (3.75,106.085) (4,110.775) (4.25,113.603) (4.5,115.252) (4.75,116.130) (5,116.484) (5.25,116.474) (5.5,116.204) (5.75,115.746) (6,115.150) (6.25,114.450) (6.5,113.674) (6.75,112.841) (7,111.966) (7.25,111.060) (7.5,110.133) (7.75,109.190) (8,108.238) (8.25,107.281) (8.5,106.321) (8.75,105.363) (9,104.408) (9.25,103.458) (9.5,102.515) (9.75,101.579) (10,100.651) (10.25,99.732) (10.5,98.823) (10.75,97.924) (11,97.035) (11.25,96.156) (11.5,95.288) (11.75,94.431) (12,93.585) (12.25,92.749) (12.5,91.924) (12.75,91.110) (13,90.306) (13.25,89.512) (13.5,88.728) (13.75,87.955) (14,87.191) (14.25,86.437) (14.5,85.693) (14.75,84.958) (15,84.232) (15.25,83.515) (15.5,82.806) (15.75,82.107) (16,81.415) (16.25,80.732) (16.5,80.057) (16.75,79.390) (17,78.731) (17.25,78.078) (17.5,77.434) (17.75,76.796) (18,76.166) (18.25,75.542) (18.5,74.925) (18.75,74.314) (19,73.710) (19.25,73.112) (19.5,72.521) (19.75,71.935) (20,71.355) (20.25,70.781) (20.5,70.212) (20.75,69.649) (21,69.091) (21.25,68.538) (21.5,67.991) (21.75,67.448) (22,66.911)};
\addplot[black, only marks, mark=square*] coordinates {(0.5,0.02) (1,0.48) (2,9.67) (3,68.93) (5,117.78) (8,108.07) (12,92.97) (16,81.01) (22,66.79)}; \addlegendentry{$L=2$}
\addplot[black, dotted, forget plot] coordinates {(0.25,0.000) (0.5,0.000) (0.75,0.000) (1,0.000) (1.25,0.001) (1.5,0.002) (1.75,0.004) (2,0.009) (2.25,0.017) (2.5,0.029) (2.75,0.046) (3,0.071) (3.25,0.105) (3.5,0.150) (3.75,0.208) (4,0.283) (4.25,0.377) (4.5,0.492) (4.75,0.633) (5,0.804) (5.25,1.007) (5.5,1.248) (5.75,1.533) (6,1.867) (6.25,2.256) (6.5,2.708) (6.75,3.232) (7,3.838) (7.25,4.537) (7.5,5.343) (7.75,6.271) (8,7.341) (8.25,8.575) (8.5,10.000) (8.75,11.647) (9,13.555) (9.25,15.770) (9.5,18.345) (9.75,21.346) (10,24.842) (10.25,28.913) (10.5,33.633) (10.75,39.065) (11,45.230) (11.25,52.088) (11.5,59.505) (11.75,67.251) (12,75.033) (12.25,82.547) (12.5,89.551) (12.75,95.893) (13,101.518) (13.25,106.440) (13.5,110.714) (13.75,114.415) (14,117.621) (14.25,120.403) (14.5,122.829) (14.75,124.951) (15,126.818) (15.25,128.469) (15.5,129.935) (15.75,131.244) (16,132.419) (16.25,133.478) (16.5,134.437) (16.75,135.309) (17,136.106) (17.25,136.836) (17.5,137.508) (17.75,138.129) (18,138.704) (18.25,139.240) (18.5,139.739) (18.75,140.206) (19,140.644) (19.25,141.056) (19.5,141.445) (19.75,141.812) (20,142.160) (20.25,142.491) (20.5,142.806) (20.75,143.106) (21,143.393) (21.25,143.667) (21.5,143.929) (21.75,144.182) (22,144.424)};
\addplot[black, only marks, mark=triangle*] coordinates {(0.5,0.00) (1,0.00) (2,0.01) (3,0.07) (5,0.81) (8,7.60) (12,75.68) (16,132.07) (22,143.43)}; \addlegendentry{$L=4$}
\end{axis}
\end{tikzpicture}
\end{center}
\caption{The $\alpha\alpha$ nuclear phase shifts $\delta_{N}$ of the BFW potential (lines) and of the fitted separable potentials (points), the screened Coulomb phase subtracted, drawn modulo $180^{\circ}$; on that scale the $L=0$ phase falls from $180^{\circ}$ through zero, the $L=2$ and $L=4$ phases rise through $90^{\circ}$ at the $^{8}$Be resonances.}
\label{fig:aaphases}
\end{figure}

\begin{table}[htbp]
\caption{\label{tab:pointcoul}Consequences of using the point Coulomb interaction with the BFW nuclear parameters in place of the folded Coulomb the potential was fitted with: the $^{8}$Be $0^{+}$ pole and the shifts of the Pauli-forbidden states of the $\alpha\alpha$ system.}
\begin{center}
\begin{tabular}{lcc}
\toprule
quantity & folded (as fitted) & point Coulomb \\
\midrule
$^{8}$Be $0^{+}$ position (keV)            & $87$ & $445$ \\
$^{8}$Be $0^{+}$ width (keV)               & $2.6$  & $78$ \\
$L=0$, $0s$ forbidden state, shift (MeV)   & ---  & $3.25$ \\
$L=0$, $1s$ forbidden state, shift (MeV)   & ---  & $1.89$ \\
$L=2$, $0d$ forbidden state, shift (MeV)   & ---  & $0.45$ \\
\bottomrule
\end{tabular}
\end{center}
\end{table}

\subsection{Pauli-forbidden-state count: a discrete Coulomb test}

The deep BFW potential carries its Pauli-forbidden states as bound states.  The bound spectrum of the original potential, bare and with the Coulomb interaction, is given in Table~\ref{tab:fbspec}. The two deep $L=0$ states are the Pauli-forbidden $0s$ and $1s$; the third is a spurious bound $^{8}$Be, expelled into the continuum once the Coulomb interaction is included. This leaves exactly the two Pauli-forbidden $L=0$ states and the one at $L=2$; $L=4$ has none.

\begin{table}[htbp]
\caption{\label{tab:fbspec}Bound spectrum of the BFW potential (MeV), bare and with the folded Coulomb interaction.}
\begin{center}
\begin{tabular}{lcc}
\toprule
state & bare & with Coulomb \\
\midrule
$L=0$, $0s$ (forbidden) & $-76.869$ & $-72.786$ \\
$L=0$, $1s$ (forbidden) & $-29.000$ & $-25.879$ \\
$L=0$, third state (spurious $^{8}$Be) & $-1.616$ & unbound \\
$L=2$, $0d$ (forbidden) & $-25.387$ & $-22.290$ \\
\bottomrule
\end{tabular}
\end{center}
\end{table}  
The Coulomb-dressed spectrum of each separable potential is computed both in the extended separable representation and by direct diagonalization of the same potential with the Coulomb force, and the two coincide; the comparison with the original potential follows. The fitted separable potentials reproduce the count of forbidden states exactly in every channel and the Coulomb-dressed positions of the $\alpha n$, the $\alpha p$ and the $\alpha\alpha$ $0d$ states, the ones the projection uses, accurately. The two $L=0$ $\alpha\alpha$ states are the documented exceptions to the energy tolerance of criterion C4: the $1s$ state lies just outside the tolerance and the deepest $0s$ state a quarter of an MeV from the original potential, a structural property of the analytic family~(\ref{eq:ffana}), which in the fits tried did not accommodate a low-momentum scattering support together with a deeply bound $0s$ state. The exceptions are acceptable because the three-body projection removes the forbidden states of the fitted representation itself (the eigenstate condition), so the deviation of these deeply bound states from their position in the original potential does not enter a three-body calculation. The adopted representation is rank~7, with the near-threshold physics carried by the support placed at the original potential's third $L=0$ bound state ($-2.283$~MeV in the short-range original potential~(\ref{eq:shortrange}), deeper than the bare $-1.616$~MeV of Table~\ref{tab:fbspec} because the subtracted term is repulsive), which with the Coulomb interaction added becomes the $^{8}$Be ground-state resonance (Sec.~\ref{sec:be8res}).

\subsection{$^{8}$Be resonances from the fitted separable potentials}
\label{sec:be8res}

\begin{table}[htbp]
\caption{\label{tab:be8}$^{8}$Be resonances from the fitted separable potentials against the experimental compilation of Ref.~\cite{Tilley2004}, relative to the $\alpha\alpha$ threshold; energies in MeV unless stated. The last column is the relative deviation of the fit from experiment, with the sign convention of Table~\ref{tab:anres}. The $0^{+}$ width entries are the bound set by the method, the screened phase at finite radius, for the fit and the original potential alike.}
\begin{center}
\begin{tabular}{lcccc}
\toprule
state & fit & original & experiment & deviation (\%) \\
\midrule
$0^{+}$: $E_{r}$ & $0.106$ & $0.087$ & $0.0918$ & $+15$ \\
\phantom{$0^{+}$:} $\Gamma$ & $<25$~keV & $<25$~keV & $5.57$~eV & --- \\
$2^{+}$: $E_{r}$, $\Gamma$ & $3.31$, $2.09$ & $3.319$, $2.110$ & $3.12$, $1.51$ & $+6$, $+38$ \\
$4^{+}$: $E_{r}$, $\Gamma$ & $12.45$, $4.12$ & $12.517$, $4.311$ & $11.44$, $3.5$ & $+9$, $+18$ \\
\bottomrule
\end{tabular}
\end{center}
\end{table}

Table~\ref{tab:be8} completes criterion C5 for this interaction; the resonances are the $\delta_{L}=90^{\circ}$ crossings of the screened-Coulomb phase at $R=80$~fm, with the width from the phase slope. The $2^{+}$ and $4^{+}$ lie above the experimental positions with larger widths, a property of the BFW model at these energies rather than of the representation, since fit and original potential agree far more closely with each other than either does with experiment; for states as broad as these the crossing definition used here and the pole definition used for the $A=5$ resonances differ, so the width deviations from experiment carry a definitional part. The experimental $0^{+}$ width of a few eV lies five orders of magnitude below its position, beyond the reach of the screened-phase method, so only the upper bound $\Gamma<25$~keV, set by the method and not by the physics, is given. The quantitative statement is that the fitted separable potential tracks the near-threshold pole of the original potential to $18$--$19$~keV (the fit and original entries of Table~\ref{tab:be8}); this is the quantity criterion C5 actually tests, and the displacement of a pole lying $87$~keV above threshold by $19$~keV is the documented limitation of the representation in this channel. Figure~\ref{fig:aascphases} shows the Coulomb-modified phases themselves in the three waves, the counterpart of Fig.~\ref{fig:aaphases} with the Coulomb interaction present and treated exactly on both sides.

\begin{figure}[htbp]
\begin{center}
\begin{tikzpicture}
\begin{groupplot}[group style={group size=3 by 1, horizontal sep=1.4cm},
  width=0.31\textwidth, height=5.2cm, xmin=0, xmax=23, xlabel={$E_{\rm cm}$ (MeV)}, ylabel={$\delta_{SC}$ (deg)},
  tick label style={font=\scriptsize}, label style={font=\scriptsize}, title style={font=\scriptsize, yshift=-1ex}]
\nextgroupplot[title={$L=0$}]
\addplot[black, solid, forget plot] coordinates {(0.5,170.43) (0.75,158.96) (1,147.72) (1.1,143.49) (1.2,139.43) (1.3,135.52) (1.4,131.78) (1.5,128.18) (1.75,119.78) (2,112.14) (2.5,98.68) (3,87.10) (3.5,76.94) (4,67.89) (4.5,59.74) (5,52.32) (5.5,45.51) (6,39.22) (6.5,33.38) (7,27.93) (7.5,22.82) (8,18.01) (8.5,13.48) (9,9.18) (9.5,5.10) (10,1.22) (10.5,-2.48) (11,-6.01) (11.5,-9.40) (12,-12.65) (12.5,-15.77) (13,-18.77) (13.5,-21.66) (14,-24.44) (14.5,-27.14) (15,-29.74) (15.5,-32.26) (16,-34.70) (16.5,-37.07) (17,-39.37) (17.5,-41.60) (18,-43.77) (18.5,-45.89) (19,-47.94) (19.5,-49.95) (20,-51.91) (20.5,-53.81) (21,-55.68) (21.5,-57.50) (22,-59.27)};
\addplot[only marks, mark=*, mark size=1.1pt, forget plot] coordinates {(0.5,170.00) (1,147.03) (1.2,138.68) (1.4,131.00) (1.75,118.99) (2.5,97.95) (3.5,76.39) (4.5,59.33) (5.5,45.18) (6.5,33.07) (7.5,22.50) (8.5,13.15) (9.5,4.79) (10.5,-2.77) (11.5,-9.67) (12.5,-16.03) (13.5,-21.93) (14.5,-27.46) (15.5,-32.66) (16.5,-37.57) (17.5,-42.21) (18.5,-46.61) (19.5,-50.77) (20.5,-54.72) (21.5,-58.45)};
\nextgroupplot[title={$L=2$}]
\addplot[black, solid, forget plot] coordinates {(0.5,0.02) (0.75,0.14) (1,0.49) (1.1,0.73) (1.2,1.07) (1.3,1.50) (1.4,2.05) (1.5,2.76) (1.75,5.38) (2,9.82) (2.1,12.34) (2.2,15.42) (2.3,19.18) (2.4,23.73) (2.5,29.19) (2.6,35.61) (2.7,42.97) (2.8,51.07) (2.9,59.55) (3,67.95) (3.1,75.82) (3.2,82.87) (3.3,88.97) (3.4,94.11) (3.5,98.39) (3.6,101.93) (3.7,104.83) (3.8,107.22) (3.9,109.17) (4,110.77) (4.1,112.08) (4.2,113.15) (4.3,114.01) (4.4,114.70) (4.5,115.25) (5,116.48) (5.5,116.20) (6,115.15) (6.5,113.67) (7,111.97) (7.5,110.13) (8,108.24) (8.5,106.32) (9,104.41) (9.5,102.51) (10,100.65) (10.5,98.82) (11,97.03) (11.5,95.29) (12,93.58) (12.5,91.92) (13,90.31) (13.5,88.73) (14,87.19) (14.5,85.69) (15,84.23) (15.5,82.81) (16,81.42) (16.5,80.06) (17,78.73) (17.5,77.43) (18,76.17) (18.5,74.92) (19,73.71) (19.5,72.52) (20,71.35) (20.5,70.21) (21,69.09) (21.5,67.99) (22,66.91)};
\addplot[only marks, mark=*, mark size=1.1pt, forget plot] coordinates {(0.5,0.01) (1,0.56) (1.2,1.01) (1.4,1.97) (1.75,5.05) (2.1,12.15) (2.3,18.64) (2.5,28.79) (2.7,42.79) (2.9,59.43) (3.1,75.84) (3.3,89.26) (3.5,98.88) (3.7,105.19) (3.9,109.50) (4.1,112.46) (4.3,114.35) (4.5,115.53) (5.5,116.38) (6.5,113.81) (7.5,110.23) (8.5,106.39) (9.5,102.57) (10.5,98.86) (11.5,95.31) (12.5,91.91) (13.5,88.67) (14.5,85.57) (15.5,82.61) (16.5,79.77) (17.5,77.06) (18.5,74.46) (19.5,71.96) (20.5,69.58) (21.5,67.29)};
\nextgroupplot[title={$L=4$}]
\addplot[black, solid, forget plot] coordinates {(0.5,0.00) (1,0.00) (1.5,0.00) (2,0.01) (2.5,0.03) (3,0.07) (3.5,0.15) (4,0.28) (4.5,0.49) (5,0.80) (5.5,1.25) (6,1.87) (6.5,2.71) (7,3.84) (7.5,5.34) (8,7.34) (8.5,10.00) (9,13.55) (9.5,18.35) (10,24.84) (10.5,33.63) (11,45.23) (11.5,59.50) (12,75.03) (12.5,89.55) (13,101.52) (13.5,110.71) (14,117.62) (14.5,122.83) (15,126.82) (15.5,129.93) (16,132.42) (16.5,134.44) (17,136.11) (17.5,137.51) (18,138.70) (18.5,139.74) (19,140.64) (19.5,141.44) (20,142.16) (20.5,142.81) (21,143.39) (21.5,143.93) (22,144.42)};
\addplot[only marks, mark=*, mark size=1.1pt, forget plot] coordinates {(0.5,0.00) (1.5,0.00) (2.5,0.03) (3.5,0.15) (4.5,0.49) (5.5,1.24) (6.5,2.70) (7.5,5.31) (8.5,9.94) (9.5,18.24) (10.5,33.52) (11.5,59.51) (12.5,89.68) (13.5,110.75) (14.5,122.66) (15.5,129.55) (16.5,133.84) (17.5,136.72) (18.5,138.77) (19.5,140.34) (20.5,141.61) (21.5,142.68)};
\end{groupplot}
\end{tikzpicture}
\end{center}
\caption{The Coulomb-modified nuclear phase shifts $\delta_{SC}$ of the $\alpha\alpha$ system: the BFW potential with its folded Coulomb interaction (lines) and the fitted separable potentials with the same Coulomb interaction treated exactly, the Pauli-forbidden states projected (points, every second grid point). The phases are drawn continuous on the branch with $\delta_{SC}=0$ at threshold; the $L=0$ phase has already risen through $180^{\circ}$ across the narrow $^{8}$Be ground-state resonance below the first plotted energy and falls from there, the $L=2$ and $L=4$ phases rise through $90^{\circ}$ at the $2^{+}$ and $4^{+}$ resonances. Fit and original potential agree to about a degree over the whole range, the separation growing slowly toward the upper end.}
\label{fig:aascphases}
\end{figure}

\FloatBarrier
\section{Three-body consequences}
\label{sec:threebody}

A separable potential satisfying every two-body criterion can still be wrong.  The acceptance procedure therefore ends with a three-body condition: the sensitivity of a fixed set of three-body solutions, the $A=3$ and $A=6$ bound states and the neutron--deuteron eigenphases of Sec.~\ref{sec:dsign}, to each element of the two-body input. The sensitivities are differences of two energies on the same mesh, in which the mesh dependence cancels, so that a keV-level difference is meaningful where a keV in the absolute energy is not.

\subsection{Sensitivity of the three-body energies to the two-body input}

\begin{table}[htbp]
\caption{\label{tab:sens}Sensitivity of the three-body energies of the adopted potential set to each element of the two-body input which the criteria constrain: the change of the $j\le4$ triton, of $^{3}$He and of the $A=6$ energies, each a difference of two energies on the same mesh. $j\le4$ denotes the CD~Bonn calculation with all $22$ pair channels of $j\le4$ (the fourteen $np$ and the eight $nn$ channels of Table~\ref{tab:npwaves}); $^{3}$He includes the $pp$ Coulomb interaction in the $S$ and $P$ waves.}
\begin{center}
\begin{tabular}{p{7.2cm}lc}
\toprule
two-body input & system & effect \\
\midrule
asymptotic $D/S$ ratio of the $^{3}S_{1}$--$^{3}D_{1}$ representation, $3.4\%$ at equal $B_{d}$ and rank (Sec.~\ref{sec:eta}); the $^{3}$H--$^{3}$He splitting moves by $16$~keV & $^{3}$H, $^{3}$He, $^{6}$Li & $\approx90$, $73$, $90$~keV \\
$\varepsilon_{1}$ constraint at high energy, deuteron constraints unchanged (Sec.~\ref{sec:eta}) & $^{3}$H, $^{6}$Li & $\sim1$, $\sim2$~keV \\
sign of the $^{3}S_{1}$--$^{3}D_{1}$ off-diagonal component relative to the kernel (Sec.~\ref{sec:dsign}) & $^{3}$H, $^{6}$Li & $7$, $30$~keV \\
$^{1}S_{0}$ $nn$ scattering length, $10\%$ from negative-energy supports alone (Sec.~\ref{sec:1s0}) & $^{6}$He & $15$~keV \\
two accepted $^{1}S_{0}$ $nn$ representations, rank $7$ and rank $5$ (Sec.~\ref{sec:1s0}) & $^{6}$He, $^{3}$H & $1$, $1.6$~keV \\
$^{3}F_{2}$ partner wave, $5^{\circ}$ at a phase below $1.3^{\circ}$ (Sec.~\ref{sec:nnphases}) & $^{3}$H & $4$~keV \\
$^{3}G_{3}$ partner wave, $1.3^{\circ}$ at $150$~MeV (Sec.~\ref{sec:nnphases}) & $^{3}$H & $0.5$~keV \\
$^{2}P_{3/2}$ rescaled to the $^{5}$He resonance position (Sec.~\ref{sec:an-res}) & $^{6}$Li & $+0.29$~MeV \\
$\alpha N$ support variation, two accepted support sets (Sec.~\ref{sec:an-phases}) & $^{6}$Li & $52$~keV \\
\bottomrule
\end{tabular}
\end{center}
\end{table}

The adopted potential set gives $E(^{6}\mathrm{Li})=-4.134$~MeV in the isospin basis with the $\alpha N$ $S$ and $P$ waves ($N_{q}=26$, converged in the mesh) and $E(^{6}\mathrm{He})=-0.8930$~MeV with the rank-$7$ $^{1}S_{0}$ $nn$ potential ($N_{q}=12$); the $A=3$ energies are given in Table~\ref{tab:a3coul}. Table~\ref{tab:sens} collects the sensitivity of these energies to each element of the two-body input which the criteria constrain. The asymptotic $D/S$ ratio of the $^{3}S_{1}$--$^{3}D_{1}$ channel is the largest effect (Sec.~\ref{sec:eta}): it shifts the triton by about $90$~keV. It moves $^{3}$He less than the triton, so the $^{3}$H--$^{3}$He splitting is sensitive to it as well: the less bound $^{3}$He is the less sensitive of the two. The sign of the $^{3}S_{1}$--$^{3}D_{1}$ off-diagonal component relative to the kernel (Sec.~\ref{sec:dsign}) is the next effect in $^{6}$Li. The threshold of the $^{1}S_{0}$ $nn$ potential (Table~\ref{tab:s0thr}) decides the $^{6}$He energy by $15$~keV, while two accepted $nn$ representations which both reproduce it are practically indistinguishable. The partner waves of the coupled channels and the $\alpha N$ support sets enter at the few-keV and the few-tens-of-keV level, respectively. Finally, the $^{2}P_{3/2}$ sensitivity study of the $\alpha N$ section bounds the contribution of the original potential's $A=5$ resonance deficit to any three-body residual.

\subsection{A condition invisible to every two-body criterion: the sign of the coupled-channel off-diagonal component}
\label{sec:dsign}

The off-diagonal component of a coupled-channel separable potential must carry the sign which the recoupling convention of the three-body kernel assigns to the tensor coupling, the same relative sign in every channel of the set. The convention dependence of coupled-channel mixing quantities is classical: the mixing parameter differs in definition and magnitude between the Blatt--Biedenharn and Stapp parametrizations~\cite{BlattBiedenharn1952,Stapp1957}, the quoted sign of $\eta$ varies with the wave-function phase convention, and the physical sign of $\eta$ is measured through interference observables~\cite{RodningKnutson1990}. The requirement here is of a different kind: a sign inconsistency internal to a potential family, one channel against the recoupling convention shared by all others, leaves every two-body quantity of the representation itself unchanged: the phase shifts, $B_d$, $P_D$, $|\eta|$, $|\varepsilon_1|$ and unitarity are invariant under a global sign change of one component of the form-factor vector, the potential being bilinear in it. A signed comparison with the original potential, of $\varepsilon_1$ or of the off-shell measure over the coupled blocks, fixes the sign relative to that potential, but not the convention of the original potential relative to the recoupling of the kernel. Which sign is the correct one for the three-body calculation is defined only relative to the recoupling convention of the kernel, and that consistency a three-body observable tests directly.

The sign is verified by a sign-reversal test: the off-diagonal component of one coupled channel is reversed in sign, the rest of the input is left unchanged, and the $3$-MeV neutron--deuteron eigenphases and the $j\le4$ triton are recomputed. The two observables differ greatly in sensitivity. For the $^{3}S_{1}$--$^{3}D_{1}$ channel the $nd$ $^{4}P_{J}$ eigenphases respond at first order, with a change of more than $10^\circ$, while the $j\le4$ triton moves by only $7$~keV, protected by a near-complete cancellation between the $S$-wave-pair binding and the exchange terms that couple the $^{3}S_{1}$--$^{3}D_{1}$ pair to the $P$-, $D$- and $F$-wave pairs of the other partitions, each of which changes by several hundred keV. The sign of the $^{3}S_{1}$--$^{3}D_{1}$ component is fixed by the $nd$ eigenphases; the bound state does not fix it at the present accuracy.

The same test is applied to the other three coupled channels (Table~\ref{tab:signcheck}). In every channel the sign as constructed is the one the tests select, but the evidence differs by channel. For $^{3}P_{2}$--$^{3}F_{2}$ the $nd$ eigenphases decide it: the reversal displaces the quartet $^{4}P_{3/2}$ eigenphase by eight times its residual against the benchmark of Ref.~\cite{Kievsky1998}, in a wave whose potential dependence between CD~Bonn and AV14 is far smaller, and for the $T=1$ representations of this channel the reversal is also worth $23$~keV in the triton. For $^{3}D_{3}$--$^{3}G_{3}$ and $^{3}F_{4}$--$^{3}H_{4}$ the eigenphases cannot decide, the signal lying below the potential dependence of the doublet $S$~wave or below the resolution of the test; there the sign is fixed relative to the original potential by the two-body sign-reversal test, in which the reversal changes the sign of the mixing parameter relative to the original potential while leaving every phase unchanged, the consistency of the convention of the original potential with the kernel being established by the $nd$ tests of the two lower channels. The remaining triton shifts under the reversals are confounded with the finite-rank cost of the representations, so Table~\ref{tab:signcheck} quotes them as sensitivities only.

\begin{table}[htbp]
\caption{\label{tab:signcheck}Sensitivity of the $3$-MeV $nd$ eigenphases and of the $j\le4$ triton to the sign of the off-diagonal component of each coupled channel: the largest eigenphase shift under a sign reversal, the wave in which it occurs, the triton shift, and the evidence that fixes the sign. For $^{3}P_{2}$--$^{3}F_{2}$ the discriminating wave is $^{4}P_{3/2}$, shifted by eight times its residual from the benchmark; the $^{4}P_{1/2}$ and $^{2}S_{1/2}$ shifts are not used, since their potential dependence exceeds the signal.}
\begin{center}
\begin{tabular}{lccc p{4.6cm}}
\toprule
channel & largest shift ($^{\circ}$) & in wave & triton (keV) & sign fixed by \\
\midrule
$^{3}S_{1}$--$^{3}D_{1}$ & $>10$   & $^{4}P_{J}$   & $7$   & $nd$ eigenphases ($^{4}P_{J}$) \\
$^{3}P_{2}$--$^{3}F_{2}$ & $0.35$  & $^{4}P_{1/2}$ & $+23$ & $^{3}$H sign-reversal test and $nd$ eigenphases ($^{4}P_{3/2}$) \\
$^{3}D_{3}$--$^{3}G_{3}$ & $0.30$  & $^{2}S_{1/2}$ & $-65$ & two-body sign-reversal test \\
$^{3}F_{4}$--$^{3}H_{4}$ & $0.01$  & $^{4}P_{5/2}$ & $+3$  & two-body sign-reversal test \\
\bottomrule
\end{tabular}
\end{center}
\end{table}

One further requirement follows from the neutron--deuteron calculation, and the two-body criteria do not enforce it. A separable potential must be free of spurious poles over the whole domain of pair energies which the three-body kernel samples, which extends far below the region of physical interest: in one rejected representation a spurious deep state at $-380$ to $-610$~MeV, with no spurious state above $-300$~MeV, shifted the $3$-MeV scattering phases by more than a degree while every two-body criterion, including unitarity, is satisfied.

With the adopted potential set the triton at $j\le4$ is $E(^{3}\mathrm{H})=-7.964$~MeV at $N_{q}=32$, converged in the mesh. The literature reference for this calculation, with the $np$ and $nn$ forces of the charge-dependent potential and no electromagnetic interaction, is the exact Faddeev value $-8.005$~MeV of Nogga \emph{et al.}~\cite{Nogga2003}, the $8.00$~MeV quoted with the potential itself~\cite{Machleidt2001}. Against it the separable representation is $41$~keV less bound. The mesh contributes less than a keV to this deficit (Table~\ref{tab:budget}); the remainder is the off-shell cost of the finite-rank representation, the kernel-weighted deviations of the $P$-, $D$- and $F$-wave families (between half a percent and twelve percent per channel), together with the pair waves beyond $j\le4$, which the exact calculation includes and the separable set does not. The same separable potentials give $^{3}$He with the $pp$ Coulomb included, and the $^{3}$H--$^{3}$He binding-energy difference tests the charge dependence and the Coulomb treatment together (Table~\ref{tab:a3coul}). A rank-one separable three-nucleon force fitted to the experimental triton energy, used here only as a diagnostic, closes the triton deficit and changes the splitting by less than $10$~keV when its regulator is varied by a factor of two: the deficit is a property of the potential, while the splitting is fixed by the two-body interaction and the Coulomb treatment.

\begin{table}[htbp]
\caption{\label{tab:budget}Error budget of the triton binding energy at $j\le4$ against the exact Faddeev value $-8.005$~MeV of Ref.~\cite{Nogga2003}.}

\begin{center}
\begin{tabular}{lc}
\toprule
source & contribution (keV) \\
\midrule
mesh truncation beyond $N_{q}=32$ & $<1$ \\
finite-rank off-shell cost and pair waves beyond $j\le4$ (remainder) & $\approx40$ \\
\midrule
total: deviation from Ref.~\cite{Nogga2003} & $41$ \\
\bottomrule
\end{tabular}
\end{center}
\end{table}  

\begin{table}[htbp]
\caption{\label{tab:a3coul}The $A=3$ system with the adopted potential set (MeV): binding energies at $j\le4$ and $N_{q}=32$, the $^{3}$H--$^{3}$He binding-energy difference $\Delta E$ against experiment, and the three-body-force test (the latter at $N_{q}=16$).}
\begin{center}
\begin{tabular}{p{8.3cm}cc}
\toprule
quantity & this work & experiment \\
\midrule
$E(^{3}\mathrm{H})$, $j\le4$, $N_{q}=32$                       & $-7.964$   & --- \\
$E(^{3}\mathrm{He})$, $j\le4$, $N_{q}=32$, $pp$ Coulomb ($S$, $P$ waves) & $-7.230$    & --- \\
$\Delta E$                                                    & $0.734$ & $0.764$ \\
triton deficit against experiment, closed by a rank-one three-nucleon force & $0.52$ & --- \\
change of $\Delta E$ under a factor-two regulator variation   & $<0.010$   & --- \\
\bottomrule
\end{tabular}
\end{center}
\end{table}

The adopted separable potentials are finally tested in three-body scattering. The $3$-MeV neutron--deuteron calculation built on them, at $j\le3$ with total angular momentum $J\le9/2$ in both parities ($28$ eigenphases, converged in the mesh), reproduces the known picture of modern $NN$-only interactions. The quartet $^{4}P_{J}$ eigenphases, the observables which fix the sign of the $^{3}S_{1}$--$^{3}D_{1}$ component (Sec.~\ref{sec:dsign}), agree with the benchmark values of the Argonne $v_{14}$ (AV14) potential~\cite{Wiringa1984} computed in Ref.~\cite{Kievsky1998}, and the higher partial waves follow the benchmark closely. The remaining differences are largest in the $S$-waves and in the $J^{\pi}=1/2^{\pm}$ blocks, their mixing parameters and eigenphases, above all in the doublet $^{2}S_{1/2}$, the channel correlated with the triton binding. They are attributed to the potential dependence between CD~Bonn and AV14 rather than to the representation, an attribution inferred from their pattern and not checked against a CD~Bonn benchmark at this energy. This agreement, together with the sensitivities of Table~\ref{tab:sens}, quantifies the off-shell error budget of this potential set.

\section{Summary}
\label{sec:summary}

A complete separable two-body input has been constructed and validated for momentum-space Faddeev calculations of light nuclei: CD~Bonn at $j\le4$ in all charge states, KKNN, and BFW with its folded Coulomb. The separable potentials are accepted against six criteria that separate construction errors from finite-rank costs, with the documented exceptions to the C4 energy tolerance for the two $L=0$ $\alpha\alpha$ forbidden states. The separable potentials are compared with their original potentials throughout, and with experiment where the comparison is meaningful (the resonances, the threshold and deuteron parameters, and the $A=3$ splitting), always read together with the comparison with the original potential. Where the separable and the original values coincide, one column stands for both.

 The $NN$ separable potentials reproduce the phase shifts and mixing parameters of CD~Bonn very well in every wave and charge state below $100$~MeV, with deviations of up to about two degrees above it, mostly in the partner waves and mixing parameters of the coupled channels; the deuteron binding energy is reproduced accurately and its asymptotic $D/S$ ratio and $D$-state probability to the values of the original potential; the scattering lengths of the three charge states are reproduced almost exactly; and the off-shell measure of every channel over the kernel energies is reported (Fig.~\ref{fig:c1nn}). The $\alpha N$ and $\alpha\alpha$ separable potentials reproduce the phases of their original potentials very well, the $A=5$ resonances to the method precision, the near-threshold $^{8}$Be pole to $18$--$19$~keV, and the Coulomb-dressed Pauli-forbidden states to a few keV, with the documented exceptions. The three-body sensitivities show which criteria matter. The asymptotic $D/S$ ratio, a ratio of residues, is the most sensitive coupled-channel test of the set: a representation that reproduces the deuteron binding energy equally well but misses the ratio by $3.4\%$ has a seven- to tenfold larger off-shell measure at the same rank, worth $\approx90$~keV in the triton, the same as in $^{6}$Li. The threshold region is not protected by supports at the negative energies the Faddeev kernel samples: the effective-range parameters must be recomputed from every fitted separable potential, a $10\%$ deviation of the $^{1}S_{0}$ $nn$ scattering length being worth $15$~keV in $^{6}$He. A phenomenological cluster potential is inseparable from the Coulomb interaction it was fitted with: for BFW the point-Coulomb substitution misplaces the $^{8}$Be ground-state resonance by $0.36$~MeV and shifts the Pauli-forbidden states by MeV. Finally, every bilinear two-body criterion is insensitive to the sign of a coupled-channel component relative to the three-body recoupling convention. The neutron--deuteron $^{4}P_{J}$ eigenphases fix that sign at first order, while the triton moves by only $7$~keV; the $nd$ test fixes the sign of $^{3}P_{2}$--$^{3}F_{2}$ as well, and the two remaining coupled channels rest on the two-body sign-reversal test. Three-body scattering must be part of the validation of two-body separable potentials.

This validation provides the computed difference from a fixed, published original potential over the kernel-sampled domain, and the three-body sensitivity to each element of the two-body input which the criteria constrain. With it, a Faddeev calculation with separable input carries a stated two-body error budget.

\bibliographystyle{unsrt}
{\small
\bibliography{refs}
}
\end{document}